\documentclass[aps,amsmath,amssymb,reprint,showkeys,superscriptaddress]{revtex4-2}

\usepackage[dvipsnames]{xcolor}
\usepackage{graphicx}
\usepackage{hyperref}
\usepackage[utf8]{inputenc}
\usepackage[T1]{fontenc}
\usepackage{physics}
\usepackage{mathtools}
\usepackage{siunitx}
\usepackage{bm}
\usepackage{dcolumn}
\usepackage{appendix}
\usepackage{lipsum}
\usepackage{multirow}
\usepackage{chemformula}

\usepackage[normalem]{ulem}

\newcommand{\Hope}[1]{\textcolor{red}{#1}}

\begin{document}
\title{Machine-learning approach for the phase stability and mechanical properties of disordered alloys at finite temperature}

\author{Rutchapon Hunkao}
\email{hunkaor@tcd.ie}
\affiliation{School of Physics, AMBER and CRANN Institute, Trinity College, Dublin 2, Ireland}
\author{Urvesh Patil}
\affiliation{School of Physics, AMBER and CRANN Institute, Trinity College, Dublin 2, Ireland}
\author{S.~Sanvito}
\affiliation{School of Physics, AMBER and CRANN Institute, Trinity College, Dublin 2, Ireland}
\date{\today}

\begin{abstract}
The prediction of stable alloys forming solid-state solutions across large portions of the composition 
space is a serious theoretical challenge, since one has to evaluate the Gibbs free energy, including 
both configurational and vibrational contributions. This requires an energy theory capable of extremely 
high throughput. By taking the Ni-Pd system as prototype, we construct an efficient Jacobi-Legendre 
machine-learning potential based on density-functional-theory data, which provides accurate energies 
and forces across the entire composition space. Based on a cluster expansion up to three-body terms
and only 873 trainable parameters, this allows us to compute the partition function by directly integrating 
all accessible microstates, differing for composition, atomic configuration and thermal agitation. We confirm 
that Ni and Pd are fully miscible, forming an {\it fcc} solid-state solution. This is only metastable at room 
temperature, while becomes thermodynamically stable at around 600~K, with the stability achieved first at 
the Pd-rich end of the composition range. Interestingly, entropy and heat capacity analysis reveal a 
competition between the solid-state solution and two intermetallic phases with long-period L1$_0$ structure for 
NiPd and NiPd$_3$. All in all, our approach offers a powerful and high-throughput workflow for the study 
of disordered alloys, an approach that can be extended to multi-component systems such as high-entropy
alloys.
\end{abstract}


\maketitle

\section{Introduction}

Disordered solid-state solutions, often found in high-entropy alloys (HEAs) \cite{George2019,Wang2023,Yang2025}, 
are at the core of many structural and functional materials. Their properties and performance are often governed 
by subtle variations in composition, atomic arrangement, and phase compositions, factors that vary across different 
temperature and pressure conditions \cite{Wang2025}. Predicting the properties of these alloys from first principles 
remains a significant challenge due to the complexity of their thermodynamic landscape and the need to consider 
large configuration spaces. While density functional theory (DFT) provides a general and reliable approach for the 
calculation of formation energies and zero-temperature phase diagrams, it becomes computationally prohibitive for 
disordered solid-state solutions, where millions of distinct atomic configurations may contribute to the system behaviour 
at finite temperature.

At zero temperature, the phase stability of an alloy is commonly assessed by using the convex hull construction, which 
compares the formation energies of all known and some hypothetical phases at fixed compositions. Since the possible 
hypothetical phases are in principle unlimited, one typically designs a sampling strategies to explore the low-energy
configurations. A phase will be considered thermodynamically stable, if it lies on the convex-hull boundary. This means 
that it has formation energy lower than any linear combination of those of other phases, namely it is stable against
any possible decomposition. In contrast, phases that lie above the convex-hull boundary are either metastable 
or unstable, since they are energetically favourable to decompose into other stable phases. This approach has been 
widely used in computational materials discovery, particularly in high-throughput DFT studies to screen stable compounds 
across the configuration and composition space \cite{Curtarolo2013, material_project}. However, since the convex hull 
approach considers only the enthalpic contribution to the Gibbs free energy of ordered structures at 0~K, it cannot directly 
describe finite-temperature stability, phase diagrams or disordered phases. These, in fact, require the calculation of the 
vibrational and configurational entropy, which are often critical for stabilizing disordered solid-state-solution phases, 
especially in HEAs \cite{Ma2015}.

To model solid-state-solution phases in disordered alloys, a widely adopted method is the special quasirandom 
structure (SQS) approach \cite{Wei1990}, where a large-enough single supercell is constructed to replicate the 
pairwise correlations of a fully random alloy. Then, the properties of the SQS structure are typically evaluated by 
DFT. Despite its utility, the SQS captures only one representative configuration and cannot account for fluctuations 
between atomic arrangements or the presence of short-range order. Furthermore, many studies combine the SQS 
with the ideal configurational entropy approximation (e.g. \cite{Jiang2004,JianG2009,Rogal2017,Chen2023,Hatzenbichler2023}). 
In this case, one replaces the actual configuration entropy with the ideal entropy of mixing, 
$S_\mathrm{conf}=k_\mathrm{B} \sum_i x_i \ln{x_i}$, where $x_i$ is the atomic faction of each element and $k_\mathrm{B}$ 
is the Boltzmann constant. While this approximation provides a convenient estimate of entropy, it only represents an 
upper bound, it assumes complete chemical disorder, and often overestimates the true entropy 
\cite{He2016,Zhang2023,Zhang2024}. As a result, the accuracy of phase-stability predictions based on this approach 
is limited.

A more rigorous phase-stability analysis at finite temperatures requires explicit sampling over the entire 
configuration space. The free energy must be evaluated as a thermal average over all relevant degrees of 
freedom, including configurational, electronic, magnetic and vibrational contributions \cite{Ikeda2019}.
Although some of these effects are often neglected or treated separately, the configurational complexity of 
disordered alloys poses a significant computational challenge due to the combinatorial explosion of the possible 
atomic arrangements. To make such calculations computationally feasible, effective models remain the most
practical solution, with the cluster expansion (CE) method \cite{Kikuchi1951,Sanchez1984,vandeWalle2002} 
being the most widely used approach. CE maps the total energy of an alloy onto an effective Hamiltonian expressed 
as a sum over effective body-ordered cluster interactions (ECIs). These are typically fitted to a DFT training set and, 
once trained, the CE model allows the rapid energy evaluation of thousands of configurations. When CE is combined with 
statistical sampling methods, it becomes a powerful tool to perform the averages required by the various thermodynamical 
quantities at a moderate computational cost. For instance, Monte Carlo sampling is often used in conjunction with 
the CE \cite{vandeWalle2002B} to compute temperature-dependent free energy via ensemble averaging \cite{Nataraj2021}. 
Wang-Landau sampling \cite{Wang2001} offers an alternative by directly estimating the configurational density of 
states (cDOS), from which the temperature-dependent free energy can also be derived \cite{Zhang2024}.
Despite its strengths, CE has notable limitations. In fact, it typically assumes a fixed parent lattice and cannot 
capture local atomic relaxations, which are often significant in disordered alloys. Although CE can be trained on 
relaxed structures, its accuracy deteriorates as the degree of local atomic relaxation increases, due to reduced 
sparsity and increased complexity of the energy landscape \cite{Nguyen2017}. Moreover, the number of clusters 
required to accurately fit the energy increases rapidly with the number of chemical species, making the model less 
scalable and substantially increasing the volume of DFT training data needed \cite{Nataraj2021B}.

To overcome these limitations, machine learning interatomic potentials (MLIPs) have emerged as a powerful 
alternative. Unlike CE, MLIPs map the potential energy surface to a continuous function of the atomic positions 
and chemical environments. This enables the efficient computation of the energy derivatives, namely forces 
and virial stresses. Thus, MLIPs can perform local relaxation and the calculation of dynamical properties, 
features particularly useful for evaluating temperature-dependent quantities such as the vibrational free energy.
Recently developed MLIPs based on neural networks \cite{NNP1}, Gaussian approximation \cite{GAP1}, spectral 
neighbor analysis \cite{SNAP1} and moment tensor \cite{MTP1}, have shown success across a broad range of 
material systems. In particular, they have been effectively applied 
to multicomponent alloys, demonstrating accurate predictions of the energetics and the thermodynamic behavior
\cite{Kostiuchenko2019,Marchand2020,Rosenbrock2021,Tolborg2023,Lin2024,Mazitov2024,Song2024,Zhu2024}.
However, despite their success, many existing MLIPs suffer from several limitations. In general MLIPs require a
large and diverse training datasets, and they can be computationally expensive to train \cite{Zuo2020}. Furthermore,
they often involve a large number of hyperparameters, the tuning of which can be non-trivial and significantly affect 
the model performance. Other issues are related to their transferability to atomic environments outside their training 
domain. Finally, many MLIPs operate as ``black-box'' models, offering limited physical interpretability, a factor 
that can hinder the understanding of the underlying atomic interactions, in particular in disordered alloys.

In this work, we employ the Jacobi-Legendre potential (JLP) \cite{JLP}, which provides a physically motivated 
and mathematically compact foundation for constructing MLIPs~\cite{Domina2025}. The JLP is 
a cluster expansion, where the different body-order terms are written over a Jacobi-Legendre polynomial 
basis set. Lower body-order contributions are directly interpretable (e.g. the two-body term appears as a 
standard pairwise interaction) and the model is linear in the clusters, so that it can be trained as a simple 
regression. As a consequence, when compared to many existing MLIPs, a JLP usually features a small and 
systematically controllable set of parameters, reducing the training complexity while maintaining high 
accuracy and strong transferability.
With these advantages, we propose an alternative approach for calculating the Gibbs free energy, as a 
function of composition and temperature. Working with a vast configurational space, we explicitly incorporate 
both local atomic relaxations and vibrational free energy contributions, going well beyond the limitations of 
conventional CE methods. Our scheme is applied to Ni-Pd binary alloys, a magnetic system, which crystallizes
as an {\it fcc} solid-state solution with complete miscibility across the composition range~\cite{Nash1984}. 

In what follows, we present our complete workflow, including strategies for the training set generation, 
hyperparameter selection, and model optimization. We demonstrate that a model trained on small 
supercells, containing up to 8 atoms, can generalize effectively to larger supercells of 16 atoms, achieving 
relatively small errors compared to the DFT reference for both relaxed total energies and vibrational free energy. 
Then, the model is applied to 24-atom supercells, for which we perform 99,268 full structural relaxations, 
followed by over 14 million force evaluations for computing the vibrational free energies. This enables us to evaluate 
the Gibbs free energy of the disordered Ni-Pd system across the full range of compositions and over a wide 
temperature range. Moreover, the model inherently allows for the direct estimation of the mechanical properties, 
which are here obtained by computing the elastic tensor. Finally, we discuss a strategy for extending our 
framework to more complex disordered alloy systems such as HEAs.

%%%%%%%%%%%%%%%%%%%%%%%%%%%%%%%%%%%%%%%%%%%%%%%%%%%%%
%%%%%%%%%%%%%%%%%%%%%%%%%%%%%%%%%%%%%%%%%%%%%%%%%%%%%

\section{Methods}\label{sec:Methods}

\begin{figure*}[ht!]
    \centering
    \includegraphics[width=0.9\textwidth]{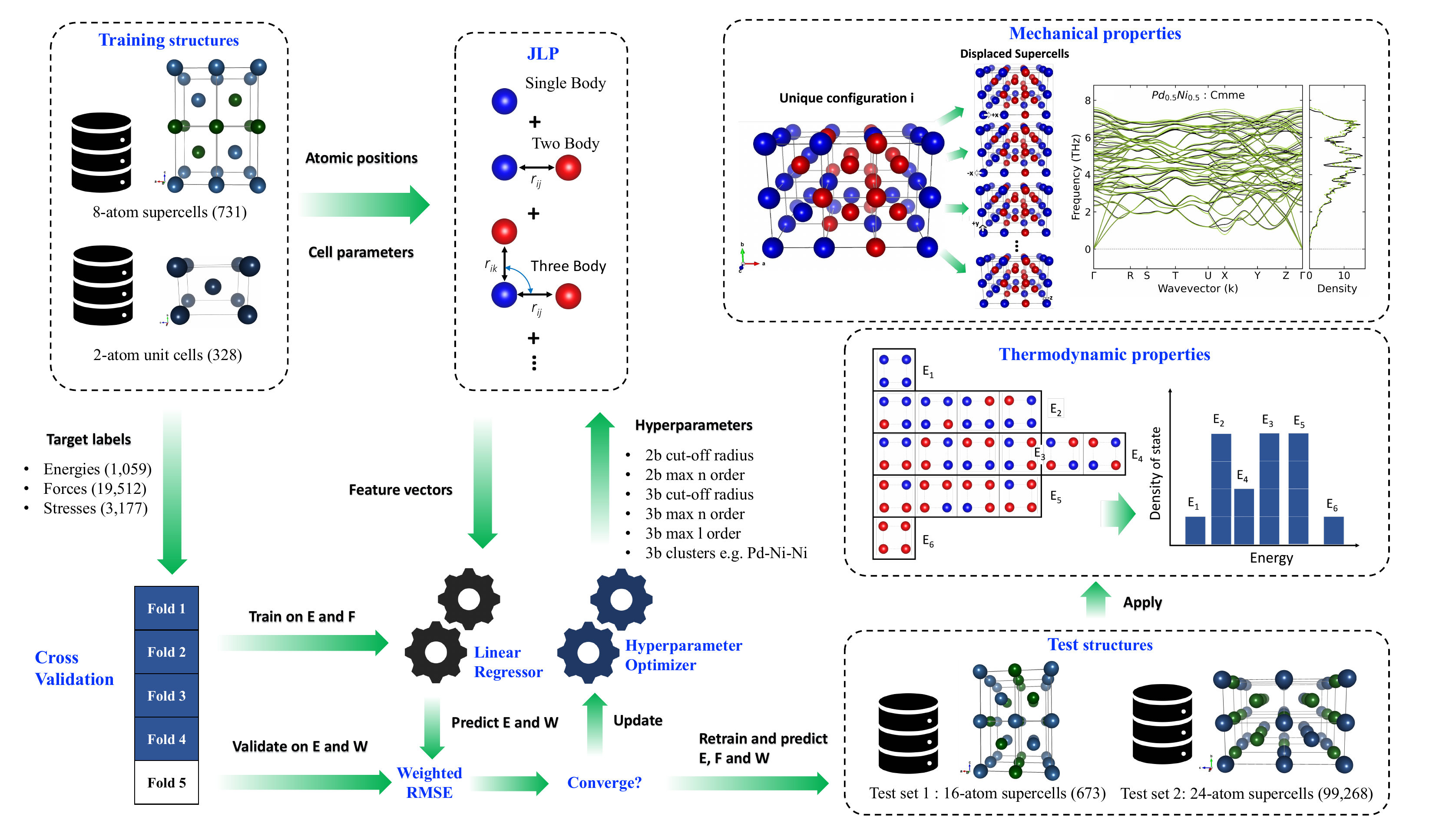}
    \caption{Scheme of the workflow proposed in this work. This comprises the creation of a training set
    for the JLP, the fit of the model and the associated optimization of the hyperparameters. Then, the
    model is deployed for the calculation of finite-temperature thermodynamical and mechanical properties
    of the Ni-Pd binary system. Here 2b (3b) indicates attributes concerning the 2-body (three-body) order
    of the JLP.}
    \label{fig:workflow}
\end{figure*}
An overview of the workflow proposed here is schematically presented in Fig.~\ref{fig:workflow}. This comprises
the JLP training, hyperparameter optimization and the prediction workflow for both thermodynamic and mechanical 
properties. These steps are described in detail in the following sections. 

% -----------------------------------------------------------------------------
\subsection{Training Structures}\label{sec:train}
\begin{figure}[htp]
    \centering
    \includegraphics[width=\linewidth]{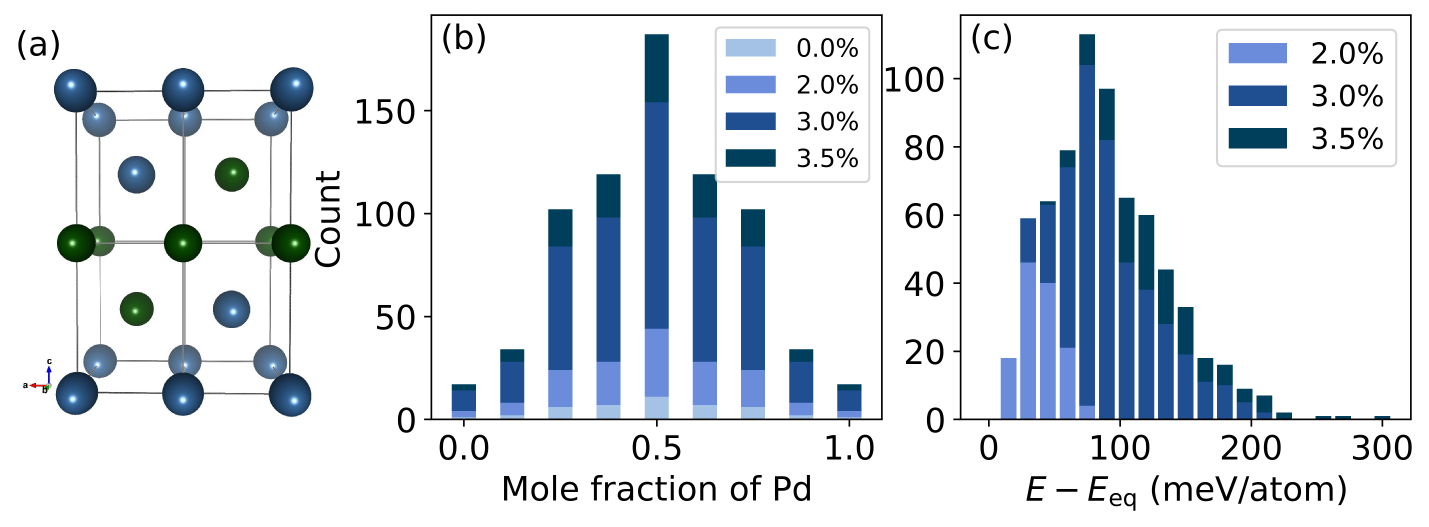}
    \caption{Composition of the training subset {\bf 1}. (a) The crystal structure of an example configuration, 
    a 2$\times$1$\times$2 {\it bct} supercell of Ni\textsubscript{0.5}Pd\textsubscript{0.5}. Blue spheres are for Pd and green
    for Ni. (b) Distribution of the compounds over the Pd mole fraction (731 structures in total). Each color on the stack 
    histograms represents the maximum value of the random perturbation applied to the configurations. (c) Energy distribution 
    of the 688 out-of-equilibrium structures relatively to their equilibrium counterparts.}
    \label{fig:supercell_train_dist}
\end{figure}
In order to train the model on the potential energy surface around equilibrium, various out-of-equilibrium 
structures have been sampled. We start with the experimental unit cells of {\it fcc} Pd and Ni, which have 
lattice constants of 0.3890~nm and 0.3524~nm, respectively~\cite{Nash1984}. Then, 43 unique (irreducible 
site-occupancy) structures made of 2$\times$2$\times$1 and 2$\times$1$\times$2 body-centered tetragonal 
({\it bct}) supercells (8 atoms), derived from the original {\it fcc} primitive cells, are identified using the Atomic 
Simulation Environment (ASE) \cite{ASE} [see Fig.~\ref{fig:supercell_train_dist}(a) for an example]. These cover 
nine possible different compositions of the Ni$_x$Pd$_{1-x}$ alloy, namely $x=0, 0.125, 0.25, 0.375, 0.5, 
0.625, 0.75, 0.875, 1$, and their distribution is shown in Fig.~\ref{fig:supercell_train_dist}(b).
Such structures are then fully relaxed by DFT (see Section \ref{sec:DFT}), and uniform random perturbations 
are applied to all the atomic positions to generate additional out-of-equilibrium structures.  For each structure, the 
position of atom $i$, ${\bm{r}_i}^\prime$, is chosen accordingly to,
\begin{equation}
     {\bm{r}_i}^\prime = \bm{r}_i + \delta_a \bm{a} + \delta_b \bm{b} + \delta_c \bm{c}\:,
\end{equation}
where $\bm{r}_i$ is the equilibrium position and $\bm{a}$, $\bm{b}$ and $\bm{c}$ are the  lattice vectors of 
the corresponding structure.
The perturbed atomic positions are obtained by choosing $\delta_a$, $\delta_b$ and $\delta_c$ randomly
within the range $\left[-\delta,\, \delta\right)$. Here we consider random displacements of 2\% ($\delta=0.02$), 
3\% ($\delta=0.03$) and 3.5\% ($\delta=0.035$), generating 3, 10 and 3 out-of-equilibrium structures, respectively. 
This results in a total of 688 out-of-equilibrium and 43 equilibrium (relaxed) structures, spanning a 
300meV/atom energy range, as shown in Fig.~\ref{fig:supercell_train_dist}(c). We will hereafter refer to this 
training subset as subset {\bf 1}.

\begin{figure}[htp]
    \centering
    \includegraphics[width=\linewidth]{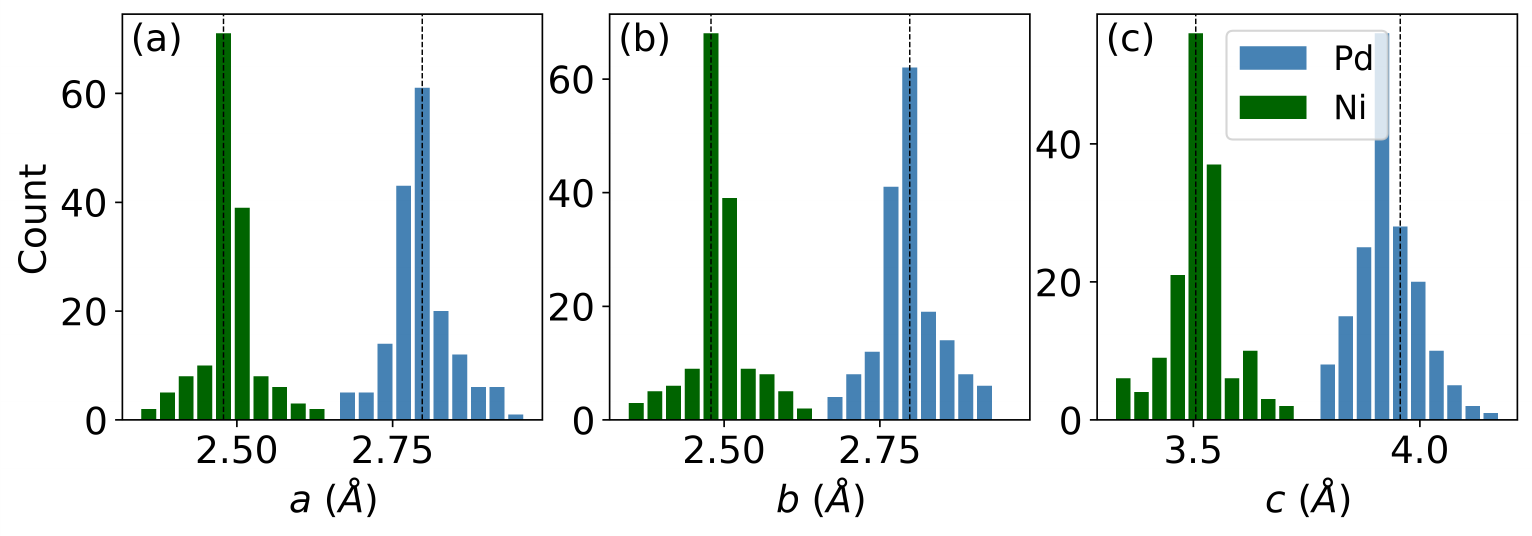}
    \caption{Distributions of the structures contained in training subset {\bf 2} (328 structures) according to
    the lattice parameters. The set includes Pd and Ni {\it bct} distorted primitive cells. Blue bars are for Pd 
    and green for Ni, with the dashed line indicating the equilibrium lattice parameters.}
    \label{fig:unitcell_train_dist}
\end{figure}
In addition, a second training subset (subset {\bf 2}) is constructed. This consists of Pd and Ni {\it bct} primitive 
cells (2 atoms per cell) with different lattice parameters. Subset {\bf 2} is designed to provide information concerning 
structures under stress, namely to improve the data needed for stress predictions, as further discussed in 
Appendix~\ref{sec:improve_elastic}. In this case, we apply random perturbations to the lattice parameters $a$, $b$ and $c$ 
(normal strains), while the cell angles and fractional atomic coordinates are kept fixed at their original values. Random 
perturbations of 5\% are applied to generate 30 distorted unit cells for both Pd and Ni. These cells are then relaxed using 
DFT, and all the relaxation trajectories are included in the dataset, resulting in a total of 328 structures spanning various 
out-of-equilibrium lattice parameter values, as shown in Fig. \ref{fig:unitcell_train_dist}. The complete training set then 
consists of both the subsets {\bf 1} and {\bf 2} for a total of 1059 structures.

% -----------------------------------------------------------------------------
\subsection{Test Structures}\label{sec:test_structures}

\begin{figure}[!ht]
    \centering
    \includegraphics[width=0.75\linewidth]{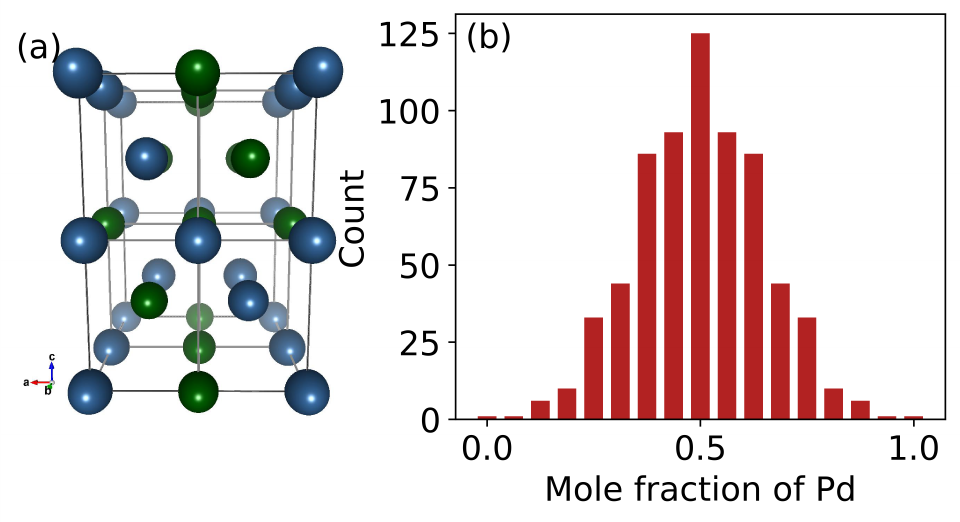}
    \caption{Summary of the model test set. (a) Crystal structure of an example configuration for the 2$\times$2$\times$2 
    {\it bct} supercell of Ni\textsubscript{0.5}Pd\textsubscript{0.5}. Blue spheres are for Pd and green for Ni. (b) Structures 
    distribution according to the Pd mole fraction for the 2$\times$2$\times$2 supercells contained in the 
    test set (673 structures).} 
    \label{fig:test_dist}
\end{figure}
The test set for the model consists of supercells larger than those contained in the training set, a feature that allows us to
explore a composition range finer than that available in the training set. Using a tree-search algorithm for irreducible 
site-occupancy configurations \cite{Lian2022}, we generate 673 unique structures for 2$\times$2$\times$2 {\it bct} supercells
(16 atoms), and 99,268 structures for 3$\times$2$\times$2 supercells (24 atoms) [see Fig. \ref{fig:test_dist}(a) for an 
example]. For each structure, the multiplicity (the number of symmetry-equivalent configurations) is also stored, since
it will be used in the calculation of the partition function, Section~\ref{sec:phase_stability}.
In the following, we refer to the 16-atom supercells as test set {\bf 1}, and the 24-atom supercells as test set {\bf 2}.
DFT calculations are then performed for test set {\bf 1} to provide reference energies, forces, and stresses, as will be 
discussed in Section~\ref{sec:model_validation}.
The 16-atom cells allow us to explore Ni$_x$Pd$_{1-x}$ alloys with $x$ varying in units of $x=1/16$, while the
24-atom cells further reduce it to 1/24. This has to be compared with the training set, where $x$ has a 1/8 resolution,
see Fig.~\ref{fig:supercell_train_dist}(b). It is also important to note that the test structures are not relaxed, and that the 
initial lattice parameters are selected by interpolation from the DFT results (see Fig.~\ref{fig:test_initial_lats} in the Appendix). 
While some structures may become symmetrically equivalent to those in the training set upon relaxation, their relaxation 
trajectories are not explicitly included in any part of the training sets.

% -----------------------------------------------------------------------------
\subsection{Machine Learning Model}\label{sec:ML_model}
Throughout this work we employ the JLP for the evaluation of energy, forces and stress tensor. For completeness, 
we will provide here a brief summary of the main concepts, with the complete treatment available in Reference \cite{JLP}. 
In the JL formalism the short-range contribution to the total energy is written in terms of a multi-body expansion,
\begin{equation}
    E_\text{total} = E_1+ E_2 + E_3 + ...\text{ },
\end{equation}
with $E_n$ representing the $n$-body ($n$B) energy term. The single-body contribution, $E_1$, is just an 
energy offset, which differentiates each atomic species present in the system. The 2B term can be described as 
the sum of pairwise interactions, $v_{Z_i Z_j}^{(2)}$, between atom $i$ and the atoms $j$ located in the 
neighborhood of $i$ within a cut-off radius $r_\mathrm{cut}$, namely
\begin{equation}
    E_2 = \sum_i^N\sum_{j \neq i} v_{Z_i Z_j}^{(2)}(r_{ji})\:,
    \label{eq:2b1}
\end{equation}
where $Z_i$ is the atomic number of the atom centered at the position $\bm{r}_i$, $r_{ji} = \left| \bm{r}_j - \bm{r}_i \right|$, 
and $N$ is the total number of atoms in the system. 

The pairwise potentials, $v_{Z_i Z_j}^{(2)}(r_{ji})$, are then expanded over Jacobi polynomials as
\begin{equation}
v_{Z_i Z_j}^{(2)}(r) = \sum_{n=1}^{n_\mathrm{max}} a_n^{Z_i Z_j}\tilde{P}_{n}^{(\alpha, \beta)}\left(\cos\left( \pi \frac{r}{r_\mathrm{cut}}\right) \right)\:.
    \label{eq:2b2}
\end{equation}
In Eq.~(\ref{eq:2b2}) we have introduced the vanishing-Jacobi polynomials, $\tilde{P}_{n}^{(\alpha, \beta)}$, which 
are defined from the standard Jacobi polynomials, $P_{n}^{(\alpha, \beta)}$, to vanish at the cut-off radius, namely
\begin{equation}
    \tilde{P}_{n}^{(\alpha, \beta)}(x) = P_{n}^{(\alpha, \beta)}(x) - \tilde{P}_{n}^{(\alpha, \beta)}(-1),
\end{equation}
for $-1\leq x \leq 1$ and $n \geq 1$. The parameter $\alpha$ and $\beta$ define the polynomial type, while $n$
is their order, so that $n_\mathrm{max}$ in Eq.~(\ref{eq:2b2}) is the maximum order in the 2B expansion.

Moving with the body-order expansion, the 3B contribution to the total energy, $E_3$, is composed 
of potentials depending on the coordinates of three atoms, $\bm{r}_{i}$, $\bm{r}_{j}$ and $\bm{r}_{k}$. 
These define a triangle, which in turn can be uniquely described by two distances and an angle. Thus, 
if the $i$-th atom sits at the vertex, the cluster is defined by the distances $r_{ji}=|\bm{r}_{i}-\bm{r}_{j}|$, 
$r_{ki}$ and the angle $s_{jki}=\hat{\bm{r}}_{ji} \cdot \hat{\bm{r}}_{ki}$, so that the 3B total energy writes
\begin{equation}
    E_3 = \sum_i^N \sum_{(j,k)_i} v_{Z_j Z_k Z_i}^{(3)} \left( r_{ji}, r_{ki}, s_{jki} \right)\:,
\end{equation}
where again $Z_i$ labels the chemical identity of atom $i$. Following the same logic of the 2B potentials,
$v_{Z_j Z_k Z_i}^{(3)}$ can be written by expressing the dependance on distances through Jacobi polynomials 
and that on the angle through Legendre polynomials, $P_l$, to obtain
\begin{equation}
v_{Z_j Z_k Z_i}^{(3)}  = \sum_{n_1, n_2=2}^{n_\mathrm{max}} \sum_{l=0}^{l_\mathrm{max}} 
a_{n_1 n_2 l}^{Z_j Z_k Z_i} 
    \Bar{P}_{n_1jl}^{(\alpha, \beta)}
    \Bar{P}_{n_2ki}^{(\alpha, \beta)}
    P_{l}^{jki},
    \label{eq:3b}
\end{equation}
where $P_{l}^{jki}=P_{l}(s_{jki})$ and $\Bar{P}_{n_1jl}^{(\alpha, \beta)}=\Bar{P}_{n_1}^{(\alpha, \beta)}(r_{ji})$.
Note that the Legendre polynomials are a special case of the Jabobi's ones obtained by taking $\alpha=\beta=0$,
and that in Eq.~(\ref{eq:3b}) we have introduced the double-vanishing Jacobi polynomials,
\begin{equation}
    \Bar{P}_{n}^{(\alpha, \beta)}(x) =  \tilde{P}_{n}^{(\alpha, \beta)}(x) - \frac{\tilde{P}_{n}^{(\alpha, \beta)}(1)}{\tilde{P}_{1}^{(\alpha, \beta)}(1)} \tilde{P}_{1}^{(\alpha, \beta)}(x),
\end{equation}
for $n \geq 2$. Then, a generic $n$B contribution beyond 3B is written in a similar way by associating to the 
distances and angles defining the cluster Jacobi and Legendre polynomials, respectively. This strategy provides 
descriptors, which depend solely on the internal coordinates between atoms so to result invariant by translations 
and rotations. In this work we truncate the cluster expansion to 3B.

The entire cluster expansion is then linear in the coefficients, which can be written in a single vector $\bm{a}$.
A model is constructed by minimizing a loss function, $l$, usually written as
\begin{equation}
    l\left(\bm{X}, \bm{a} \right) =  L({E} - \bm{X}_E \bm{a}) + c_F L(\bm{F} - \bm{X}_F \bm{a})  + c_W L(\bm{W} - \bm{X}_w \bm{a})\:,
%    \|\bm{E} - \bm{X}_E \bm{a} \|_2^2 + c_F \| \bm{F} - \bm{X}_F \bm{a} \|_2^2  + c_W \| \bm{W} - \bm{X}_w \bm{a} \|_2^2.
    \label{eq:loss_model}
\end{equation}
with $L$ being a convenient metrics. In our case we take $L$ to be the $L_2$ norm, $L(\bm{x})=\|\bm{E} - \bm{x}\|_2^2$, 
which is then computed independently over the energy, forces and the components of the stress tensor. The coefficient 
$c_F$ and $c_W$ are weights 
scaling the contributions of each term and are treated as hyperparameters. The vector $\bm{E}$ includes the DFT-calculated 
energies of all configurations in the training set, and the matrix $\bm{X_E}$ contains the corresponding descriptors on each 
row. Similarly, $\bm{F}$ and $\bm{W}$ contain the $3N$ components of the forces and the 6 unique components of the stress 
tensor, respectively, for each configuration, with $\bm{X_F}$ and $\bm{X_W}$ hosting their corresponding differentiated 
descriptors.
% -----------------------------------------------------------------------------

\subsection{Hyperparameter Optimization}\label{sec:hyper_opt}
When the cluster expansion is terminated at the 3B level the model involves 12 hyperparameters, as summarized in 
Table \ref{tab:hyperparam}. All possible 2B contributions are included in the model, namely the Pd-Pd, Pd-Ni and Ni-Ni
pairwise interactions. These are determined by the parameters $(\alpha, \beta)$ defining the type of Jacobi polynomials,
by the cutoff radii, $r_\mathrm{cut}$, and by the maximum polynomial order, $n_\mathrm{max}$. 

In contrast, there are six distinct 3B clusters, namely Pd-Pd-Pd, Pd-Pd-Ni, Pd-Ni-Ni, Ni-Pd-Pd, Ni-Pd-Ni and Ni-Ni-Ni.
Since each 3B term introduces several parameters and not all the clusters are equally representative, only a subset 
of size $s=1,\dots,6$ is selected. Thus, the integer $c_{3B}$ serves as a global index enumerating all the non-empty 
subsets of the six clusters. Thus, for $s=1$, $c_{3B}=0,\dots,5$, corresponding to the individual 3B clusters. Then,
for $s=2$, we have $c_{3B}=6,\dots,20$, which map all the possible pairs of 3B clusters. Finally, for $s=6$, $c_{3B}=62$ 
corresponds to the full set of six 3B clusters. Overall, $c_{3B} \in [0,62]$ spans all the 63 non-empty 3B combinations.
The remaining hyperparameters for the 3B contribution are the same as those for the 2B, with the addition of
$l_\mathrm{max}$, which is the maximum order of the Legendre polynomials that describe the angular dependence.
Finally, the forces and stress tensor weights of the loss function, respectively $c_F$ and $c_W$, complete the
hyperparameter pool of the model.
\begin{table}[ht]
    \caption{Hyperparameters of the model and their value range used during the optimization process. 
    Parameters are associated to the 2B and 3B energy contribution and to the composition of the loss 
    function. The values corresponding to the best trial are reported in the column `best'.}
    \centering
    \begin{tabular}{l|llllll}
        & Name & Min & Max & Best & Type & Description \\
        \hline\hline
         \multirow{4}{*}{2B} & $\alpha$ & \multicolumn{2}{c}{1.00} & 1 & $\mathbb{R}$ & Jacobi poly. type \\
         & $\beta$ & \multicolumn{2}{c}{1.00} & 1 & $\mathbb{R}$ & Jacobi poly. type \\
         & $r_\mathrm{cut}$ & 4.3 & 8.0 & 6.392 & $\mathbb{R}$ & Cut-off radius \\
         & $n_\mathrm{max}$ & 6 & 12 & 7 & $\mathbb{Z}^+$ & Jacobi poly. max order \\
         \hline
         \multirow{6}{*}{3B} & $\alpha$ & \multicolumn{2}{c}{1.00} & 1 & $\mathbb{R}$ & Jacobi poly. type\\
         & $\beta$ & \multicolumn{2}{c}{1.00} & 1 & $\mathbb{R}$ & Jacobi poly. type \\
         & $r_\mathrm{cut}$ & 3.3 & 6.0 & 4.925 & $\mathbb{R}$ & Cut-off radius\\
         & $n_\mathrm{max}$ & 4 & 9 & 6 & $\mathbb{Z}^+$ & Jacobi poly. max order \\
         & $l_\mathrm{max}$ & 2 & 9 & 9 & $\mathbb{Z}^+$ & Legendre poly. max order \\
         & $c_{3B}$ & 0 & 62 & 57 & $\mathbb{Z}^+$ & 3B clusters \\
         \hline
          \multirow{2}{*}{$l$} & $c_F$ & 0.001 & 0.6 & 0.317 & $\mathbb{R}$ & Force weight\\
         & $c_W$ & \multicolumn{2}{c}{-} & - & $\mathbb{R}$ & Stress weight \\
        \hline\hline
    \end{tabular}
    \label{tab:hyperparam}
\end{table}

\begin{figure*}[!ht]
    \centering
    \includegraphics[width=0.85\linewidth]{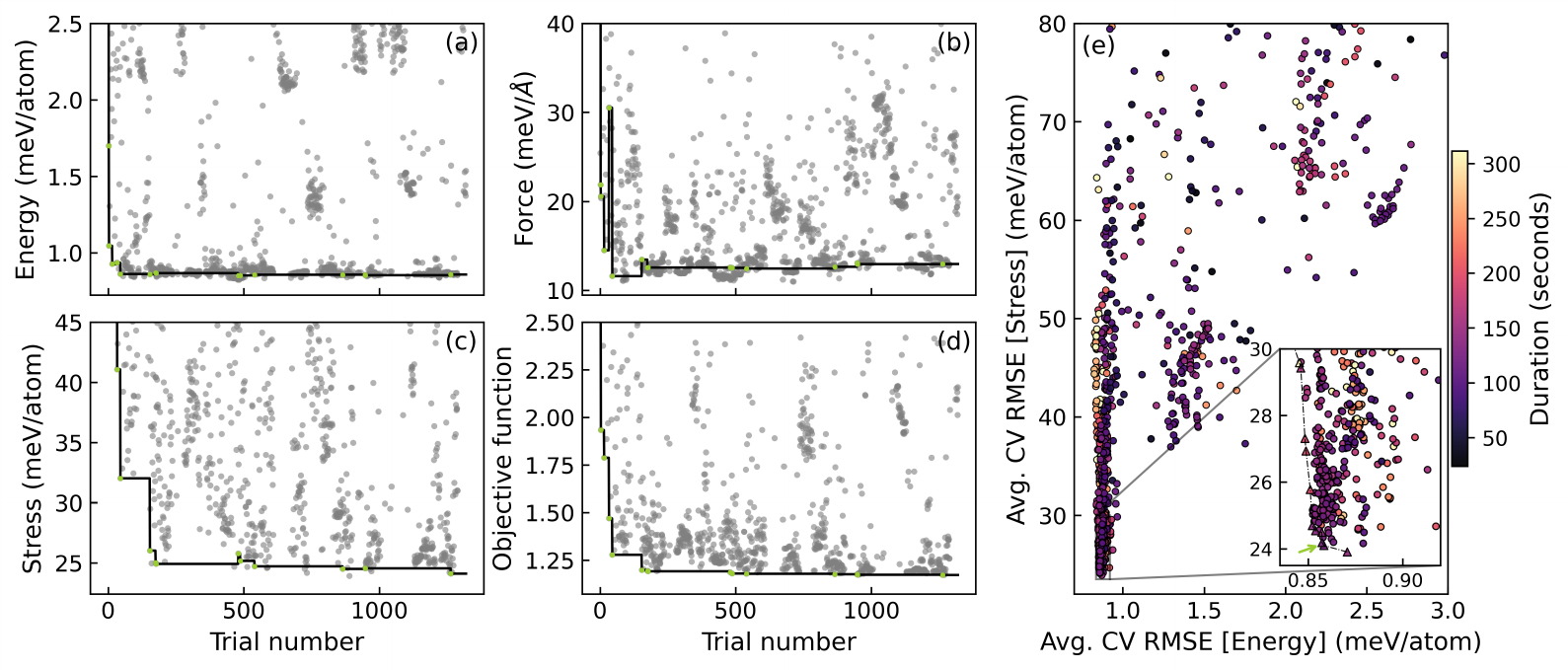}
    \caption{Hyperparameter optimization and model selection. The average cross-validation RMSE of 
    (a) energy, (b) forces, (c) stress and (d) value of the objective function are plotted as a function of the
    trail number. The black solid lines represent the convergence line, with green circles indicating trials with 
    new minimum values (according to the objective function), and grey points depicting the remaining trials. 
    In panel (e) we present the cross-validation RSME for the stress against that of the energy, with the color 
    code encoding the time taken to execute a trial, taken as a proxy for model complexity. The color bar is 
    capped at the 95th percentile to enhance visibility. The inset shows the best trial, indicated by the green 
    arrow, and other optimal points on the Pareto front. This is marked by triangles.}
    \label{fig:hyperparam}
\end{figure*}
Training is performed using a k-fold cross-validation (CV) strategy, where the model is evaluated  across multiple 
training subsets to reduce overfitting. This has been implemented within scikit-learn \cite{scikit-learn}. An objective 
function is written as a weighted combination of the errors over energy and stress, $\epsilon_{E}^i$ and 
$\epsilon_{W}^i$, as
\begin{equation}
    l_h = \bar{\epsilon}_E + c_h \bar{\epsilon}_W =  \sum_{i=1}^m \frac{\epsilon_{E}^i\left( \bm{X}_i, \bm{a}_i  \right)}{m} + c_h \sum_{i=1}^m \frac{\epsilon_{W}^i\left( \bm{X}_i, \bm{a}_i  \right)}{m}\:,
    \label{eq:loss_hyper}
\end{equation}
where $i$ indicates the $i$-th set of the test subset during the CV, and $\bar{\epsilon}$ is the average 
error over the subset. A 5-fold CV ($m=5$) is then used with the root mean square error (RMSE) 
representing the errors $\epsilon$ on each test subset. Notably, we find that omitting the stress tensor 
(set $c_W=0$) during the model fitting [see Eq.~(\ref{eq:loss_model})] and instead using it to validate 
the hyperparameters [$c_h$ term in the Eq.~(\ref{eq:loss_hyper})] is crucial, as this strategy significantly 
reduces overfitting when extending calculations to larger supercells. In this approach, the model is trained 
exclusively on energies and forces (the loss function does not include the stress component), while 
stresses are derived from the model and compared to DFT during the hyperparameter optimization through 
the weighted error term. Evaluating performance on stresses without directly fitting them acts as an implicit 
regularization mechanism, thereby improving the transferability to larger systems. The $c_h$ value has 
been heuristically set at 0.013 to achieve an appropriate balance between the accuracy over energy, 
force and stress tensor. 

The hyperparameters optimisation is then performed by searching for the lowest value of $l_h$ with 
{\sc optuna} \cite{optuna_2019}, running a tree-structured Parzen estimator (TPE) sampler. As outlined earlier, 
the search space is described in Table \ref{tab:hyperparam} with $\alpha=\beta=1$ being fixed to reduce the 
complexity of the optimization. The search was performed over 1320 iterations and the optimal point was 
found at the trial 1264. This contains 873 features and we find $c_{3B}=57$, suggesting that the Ni-Pd-Ni 
cluster is not needed. The cross-validation RMSEs are 0.86~meV/atom, 12.96~meV/\AA\ and 24.10~meV/atom 
respectively for the energy, forces components and stress tensor, respectively.

The convergence plot is shown in Fig. \ref{fig:hyperparam}(d), where the value of the objective function
is plotted against the trail number. Clearly, one can notice a rapid saturation, with the objective function 
reaching a minimum plateau after only a few hundred trials. The RMSEs on energy, forces and stress 
[see Fig. \ref{fig:hyperparam}(a), \ref{fig:hyperparam}(b) and \ref{fig:hyperparam}(c), respectively] converge 
to reasonably small values, with only a few trials presenting RMSEs smaller than that marked by the 
convergence line. This is the line joining subsequent trails presenting minimum objective function. 
Importantly, there is a trade-off between the accuracy of each quantity and the model complexity. 
This can be appreciated by looking at Fig.~\ref{fig:hyperparam}(e) where we show the RMSE of the stress
tensor against that of the energy for the optimization trials. The color code indicates the time spent to
run the model, which is taken as a proxy for the model complexity. As shown by the Pareto front, indicated
by the triangles in the inset of Fig. \ref{fig:hyperparam}(e), there is a full range of models with similar 
performance and a trade off in complexity. One then has the freedom to select the most appropriate
model depending on the specific applications. For instance, a model with the lowest energy error may be 
selected for improved energy prediction, but a more compact model (indicated by darker colors) can be 
chosen for faster computation without a significant reduction in accuracy. In this work, the model is retrained 
on the entire training set using the hyperparameters from the best trial [the one indicated by the green arrow
in the inset of Fig.~\ref{fig:hyperparam}(e)], and the results are discussed in Section~\ref{sec:model_validation}. 

% -----------------------------------------------------------------------------
\subsection{DFT and phonon calculations}\label{sec:DFT}
The Vienna ab-initio simulation package ({\sc vasp}) \cite{vasp1, vasp2, vasp3} is employed throughout this 
work to calculate the energy, forces and stresses dataset. Spin-polarized DFT with the Perdew-Burke-Ernzerhof 
(PBE) exchange correlation functional \cite{PBE} is considered. The projector augmented wave (PAW) potentials 
\cite{PAW} are used for both Pd and Ni with valance electron configuration of 4d\textsuperscript{9} 5s\textsuperscript{1}  
and 3d\textsuperscript{9} 4s\textsuperscript{1}, respectively. The cutoff energy and $k$-point spacing are 439~eV and 
0.25 \AA, respectively, values that ensure the energy convergence across the dataset. 
All the relaxations are performed until the forces on each atom are less than 0.02~eV/atom. 
Phonon calculations are performed using the finite-displacement method as implemented in the {\sc phonopy} 
package \cite{phonopy}. Since our simulation cells are constructed as $2\times2\times2$ and $3\times2\times2$ 
supercells of the {\it bct} unit cell, these are directly used for the phonon calculations with a displacement amplitude 
of 0.01~\AA. The same DFT settings as described above are used for force calculations to ensure consistency. 
For each configuration, we compute the phonon spectra on a $20\times20\times20$ $q$-point mesh and 
obtain the vibrational band structure, density of state and free energies in the 0-1500~K range in 10~K steps. 
% -----------------------------------------------------------------------------

%%%%%%%%%%%%%%%%%%%%%%%%%%%%%%%%%%%%%%%%%%%%%%%%%%%%%
%%%%%%%%%%%%%%%%%%%%%%%%%%%%%%%%%%%%%%%%%%%%%%%%%%%%%

% -----------------------------------------------------------------------------
\section{Results and discussion}\label{sec:Results}

% -----------------------------------------------------------------------------
\subsection{Model validation}\label{sec:model_validation}
Firstly, the trained model is evaluated against the training and test sets, so to provide an estimate of 
its accuracy and ability to predict the DFT energies, forces and stresses of unseen configurations.  
\begin{figure*}[ht]
    \centering
    \includegraphics[width=0.85\linewidth]{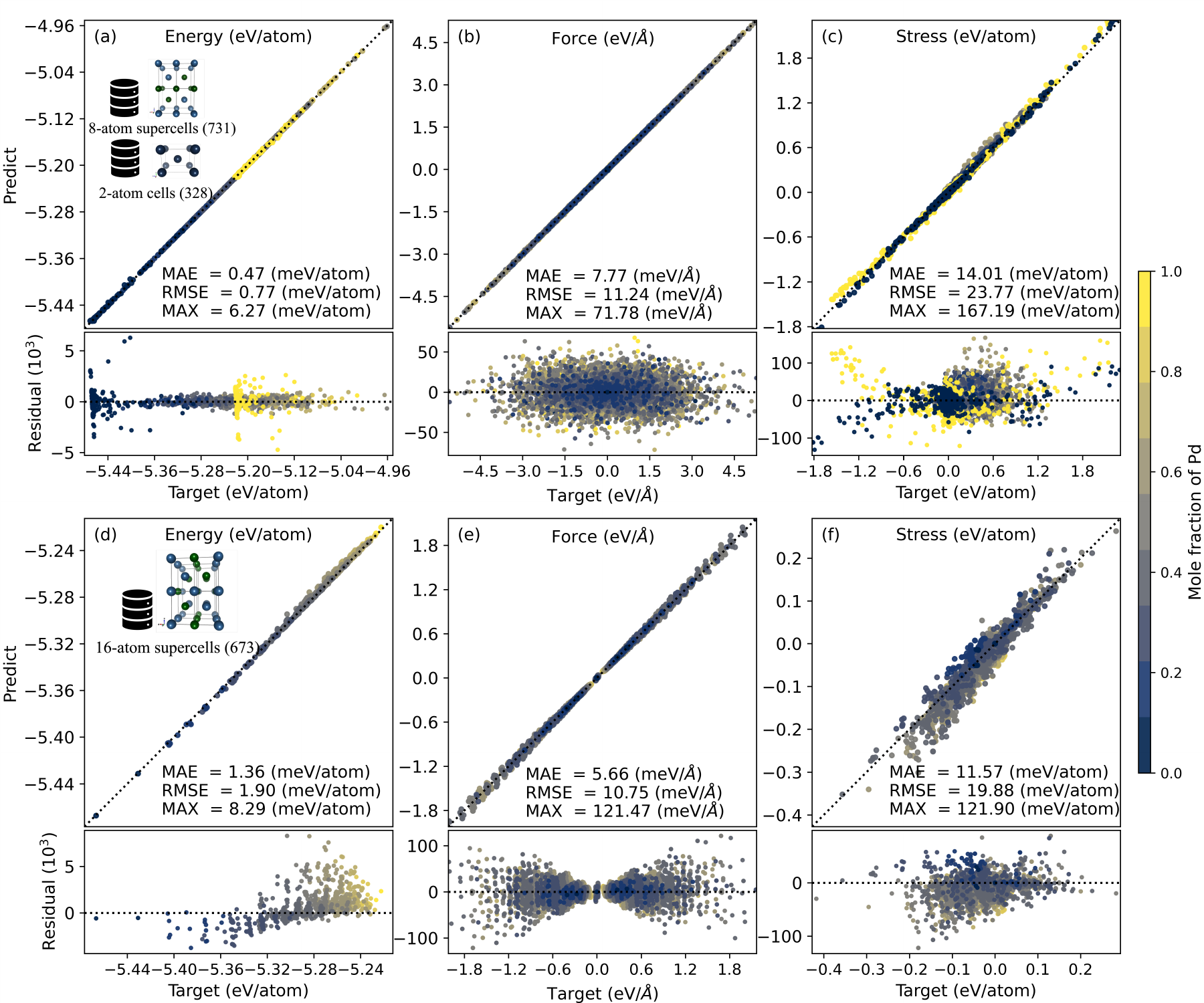}
    \caption{Parity plots for energy, forces and stress tensor computed over the training set [panels (a)-(c)] and test set {\bf 1}
    [panels (e)-(g)]. The plot compare the JLP predictions with the DFT reference values, with the dotted line indicating
    the parity line. The root mean square error (RMSE), mean absolute error (MAE), and maximum absolute error (MAX) 
    are reported in the legends. The color code encodes the Pd mole fraction of each structure. For each parity plot we
    also report a plot of the residual error.}
    \label{fig:train_err}
\end{figure*}
As illustrated in panel (a) through (c) of Fig.~\ref{fig:train_err}, the JLP fits rather well the training set, with RMSE values 
of~0.77 meV/atom for energy, 11.24~meV/$\AA$ for forces, and 23.77~eV/atom for the stress tensor. These are typical
values for the JLP~\cite{JLP} and has to be considered extremely accurate, considering that the model is linear and
constructed on less than a thousand features. A similar performance is observed on the test set {\bf 1}, Fig.~\ref{fig:train_err}(e)
throughput Fig.~\ref{fig:train_err}(g), with RMSE values of 1.9~meV/atom for energy, 10.75~meV/$\AA$ for forces and 
19.88~eV/atom for stress, suggesting that the model does not overfit the training data and indeed performs well across
a multitude of configurations and stoichiometry.

Going in more detail of the error distribution, we observe in the Fig.~\ref{fig:train_err}(c) that the large stress values 
and errors primarily originate from the training subset {\bf 2}, while smaller values are associated to the training subset 
{\bf 1}. This is by design since we aimed at improving the model accuracy across a wide stress range. As such, the 
model was not fitted directly to the stresses $c_W=0$, a strategy that ensures transferability to larger supercells, as 
discussed in Section~\ref{sec:hyper_opt}.
\begin{figure}
    \centering
    \includegraphics[width=0.9\linewidth]{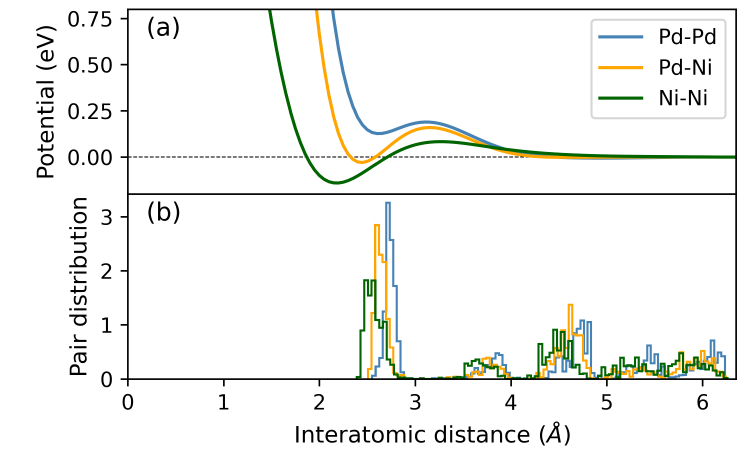}
    \caption{Reconstructed 2B potential. (a) the 2B potential for all possible clusters: Pd-Pd, Pd-Ni and Ni-Ni. (b) Atom pair 
    distributions calculated from all the relaxed structures in the training subset {\bf 1}. Note that the potentials are all repulsive
    at short distance and they tend to have a minimum at around the first neighbour peak of the relative pair-distribution function.}
    \label{fig:reconstruct_pot}
\end{figure}

A key advantage of the JLP is that the 2B and 3B potentials are fully interpretable, a feature that alone allows one to
understand whether overfitting is taking place. As an example, in Fig.~\ref{fig:reconstruct_pot} we show the reconstructed
2B potential [panel (a)] together with the corresponding pair-distribution function [panel (b)]. In general, all pairwise potentials 
(Ni-Ni, Pd-Pd and Ni-Pd) are smooth with a minimum at around the position of the nearest neighbour peak of the corresponding
pair-distribution function. This is not exactly at the minimum since it is corrected partially by the 3B contribution, and it corresponds
to an ensemble of structure and not a single one. Furthermore, our 2B potentials are all repulsive at short distance, a feature
that avoid structure collapse in relaxation and molecular dynamics simulations. This, in fact, is not a trivial property to enforce 
over a ML potential, and it is not automatically guaranteed, for instance, in GAPs, NNPs and graph-based models. Sometime
one can augment a short-range repulsive behaviour by imposing constraints, or explicit short-range repulsive 
terms \cite{Wang2019, Liu2023}. 

% -----------------------------------------------------------------------------
\subsection{Structural relaxation}\label{sec:relaxation}
Local atomic relaxations have a significant impact on the energy and phase-stability prediction of disordered 
alloys, where local atomic environments vary significantly from their ideal lattice sites due to the different ionic 
sizes of constituent elements \cite{He2018}. Neglecting relaxation can lead to substantial errors in energy 
predictions, while including lattice relaxations using MLIPs can efficiently improve the accuracy of phase-stability 
predictions, as discussed in reference \cite{Kostiuchenko2019}. 

\begin{figure}[htp]
    \centering
    \includegraphics[width=1\linewidth]{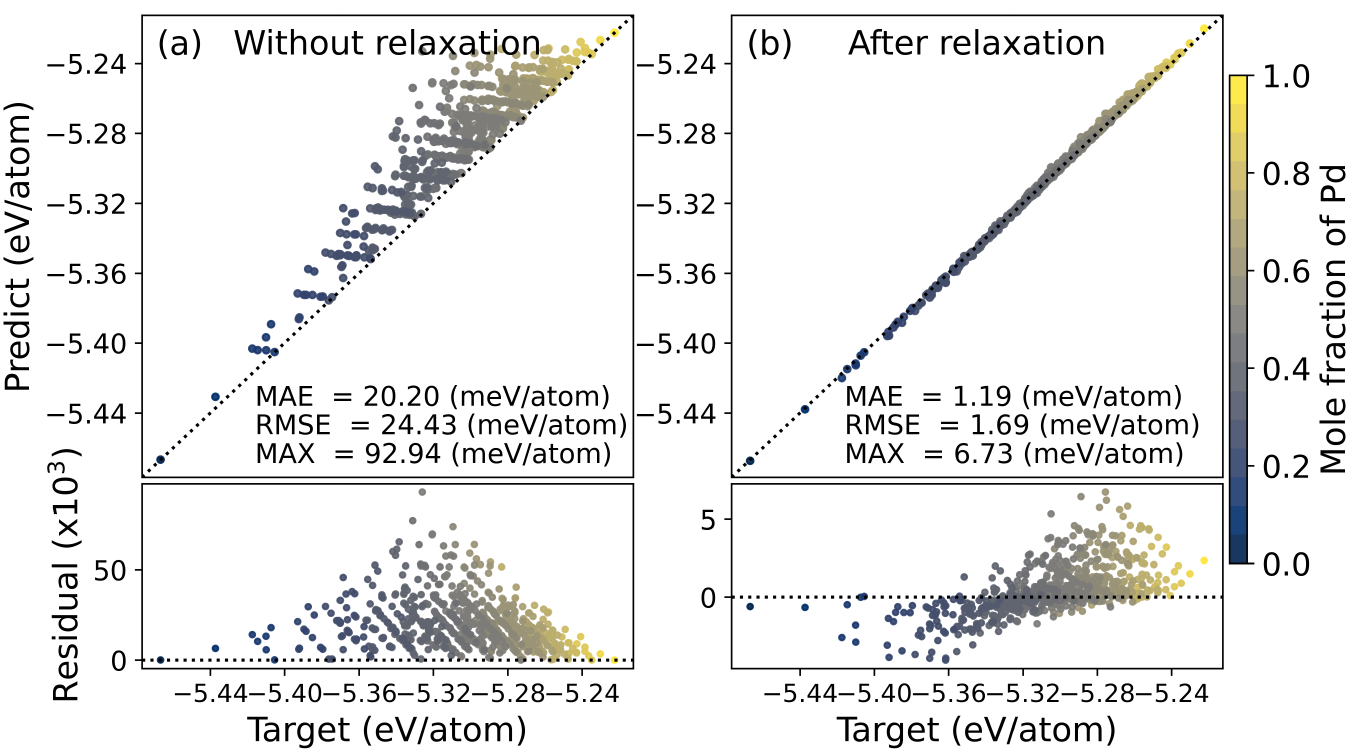}
    \caption{Comparison of JLP energy predictions relative to DFT‑relaxed energies for test set {\bf 1}.
    (a) JLP energies evaluated on unrelaxed initial structures compared with DFT‑relaxed energies. 
    (b) JLP‑relaxed energies against the DFT‑relaxed energies.}
    \label{fig:test_relax_vs_nonrelax}
\end{figure}

To demonstrate the importance of structural relaxation for our system, all structures in test set {\bf 1} are fully 
relaxed using DFT to obtain reference structures and their corresponding energies. In the context of phase-stability 
predictions, the DFT‑relaxed energies serve as the ground‑truth values, while the JLP values represent predictions.
It is worth emphasizing that many studies have assessed phase stability using directly unrelaxed energies.
However, our results show that approximating the energy of a configuration using its unrelaxed initial structure 
can lead to substantial deviations, as shown in Fig.~\ref{fig:test_relax_vs_nonrelax}(a).
In this case, the energies are systematically overestimated across all compositions, with a RMSE of 23.87 meV/atom. 
The deviation is most pronounced near equiatomic compositions, where local chemical and size mismatches are 
strongest. This substantial overestimation persists despite the fact that the initial structures were constructed to closely 
approximate their relaxed counterparts by interpolating lattice parameters from DFT results (see 
Fig.~\ref{fig:test_initial_lats} in the appendix). It is important to note that a comparable bias would also be expected 
if DFT were evaluated on the same unrelaxed initial structures, as evidenced by small error between DFT and JLP 
for these structures shown in Fig.~\ref{fig:train_err}(d). This demonstrates that, even though MLIPs are often reported 
to achieve low energy prediction errors, they can still fail to provide accurate energy predictions without proper 
structural relaxation.

In contrast, when structural relaxation is performed by interfacing the JLP with ASE results drastically improve.
The relaxation is here performed in two steps by using the Broyden–Fletcher–Goldfarb–Shanno (BFGS) algorithm 
with a force tolerance of 0.005~eV/\AA. In the first step the atomic positions are relaxed while keeping 
the unit cell fixed, and in the second step the unit cell is relaxed by fixing the atomic fractional coordinates.
As shown in Fig.~\ref{fig:test_relax_vs_nonrelax}(b), the relaxed energies are predicted with a small RMSE of 
1.6~meV/atom (in comparison with DFT-relaxed energies), and this is obtained with an extremely high computational 
efficiency, which will be discussed further in Section~\ref{sec:extension}. It should be emphasized that the JLP-relaxed 
structures and energies are obtained using JLP alone without any DFT computations, starting from unrelaxed initial 
structures (prototypes obtained by decorating ideal lattices). Further investigation reveals that the JLP-relaxed cell 
parameters (distances and angles) agree well with the DFT results, as shown in Fig.~\ref{fig:test_lattice_param}. 
Similar good agreement is also found for the atomic pair distributions and space group symmetries (up to about 
$10^{-4}$). These, in fact, are extremely close to those obtained from DFT as shown in Fig. \ref{fig:symprec_pair_dist} 
in the appendix.

% -----------------------------------------------------------------------------
\subsection{Zero-temperature convex-hull diagram}
After performing structural relaxations, we evaluate further the model predictive performance by calculating 
the formation enthalpy. Thus, we construct the zero-temperature convex-hull diagram for the Ni-Pd system, 
and compare the results to DFT data across a broad stoichiometry and structure range. Here, we use test 
set {\bf 2} (24-atom cell) to provide more detailed information across a wider range of mixing compositions, 
and also to evaluate how well the model extends to larger structures. All the 99,268 prototype structures are 
fully relaxed using JLP, and their formation energy, $E_{\alpha}^\mathrm{form}$, are calculated as
\begin{equation}
    E_{\alpha}^\mathrm{form}(\sigma) = E_{\alpha}(\sigma) -  \sum_i^N x_i^{\alpha} E_i\:,
    \label{eq:formation_energy}
\end{equation}
where $E_{\alpha}(\sigma)$ is the total energy of the $\sigma$ configuration of the multicomponent alloy, 
and $x_i^{\alpha}$ is molar composition of $i$-th atomic species (up to $N$) with lattice type $\alpha$. 
In Eq.~(\ref{eq:formation_energy}) the $E_i$'s are energies of the most stable unary phases ({\it fcc} in 
this case). A negative formation energy indicates the possible thermodynamic stability of the given structure 
at 0 K, while a positive value is for structures unstable against phase separation into the elemental structures.
Note that, in principle, the formation energy should be computed against all possible phases, which may 
include decompositions involving one or more binary compounds. In the case of the Ni-Pd system, however, 
there is only a single binary with negative $E_{\alpha}^\mathrm{form}(\sigma)$, meaning that decompositions
along elemental phases are the ones determining the convex-hull stability boundary. Furthermore, as
suggested by the experimental phase diagram, we consider here only intermetallic phases derived from the 
{\it fcc} lattice. 
\begin{figure}
    \centering
    \includegraphics[width=1\linewidth]{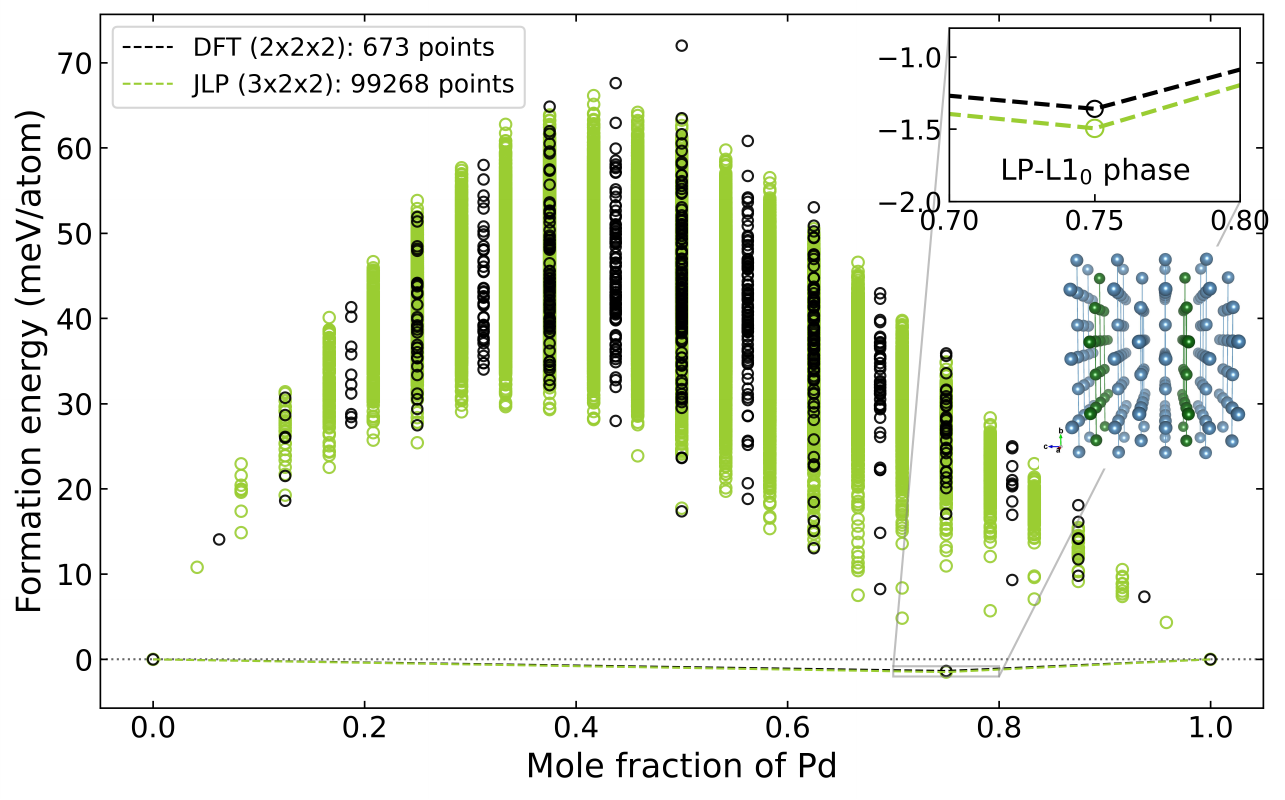}
    \caption{Zero-temperature convex-hull diagram for the Ni-Pd {\it fcc} system. We plot $E_{\alpha}^\mathrm{form}$ 
    across different compositions and structures, by either using JLP (green circles) or DFT (black circles) energies. 
    In the inset we zoom in the region where the only stable binary is present, whose crystal structure is also displayed. 
    Green spheres are for Ni and blue for Pd.}
    \label{fig:convex_hull}
\end{figure}

The computed zero-temperature convex-hull diagram is presented in Fig.~\ref{fig:convex_hull}. As one can see, the
agreement between the JLP predictions and the DFT reference data is quite strong, with energy deviations in the
region of 0.1~meV/atom. This accuracy allows one to confidently assign stability to compounds~\cite{Rossignol2024}.
The comparison is carried out over 673 structures, all constructed from 2$\times$2$\times$2 {\it bct} supercells (test 
set {\bf 1}). However, the light computational requirements of the JLP allow us to extend this analysis to 3$\times$2$\times$2 
supercells for a total of 99,268 structures (test set {\bf 2}). From this analysis it emerges that there is only a single 
stable structure across the entire composition range, namely NiPd$_{3}$. This is an intermetallic with 
$E_{\alpha}^\mathrm{form}(\sigma)\sim1.5$~eV/atom, namely just below the stability line. 
The compound has a four‑layer long‑period L1$_0$ derived ordering (LP‑L1$_0$) (see the inset), which belongs to 
the family of long‑period superlattices (LPSs). Interestingly, one‑dimensional LPSs in {\it fcc} $A_3B$ alloys, such as 
Cu$_3$Pd, Cu$_3$Al, and Ag$_3$Mg, have been both experimentally observed and theoretically shown to be 
energetically lower than conventional ordered phases (e.g., L1$_2$) at 0~K~\cite{Rosengaard1994}.
This stabilization arises from the energetics of periodically spaced antiphase boundaries and the associated 
band‑structure (Fermi‑surface nesting / charge‑density‑wave) effects in L1$_0$/L1$_2$‑derived
superlattices~\cite{Koyama1991,Sato1962}.
As we will show in the following, such structure is unlikely to be stable at room temperature or to be accessible 
by cooling from a liquid phase. However, it may be stabilised by extremely prolonged low-temperature
annealing, as demonstrated recently for equiatomic L1$_0$ FeNi~\cite{Lewis2023}. Note also that, several experiments 
\cite{Abys1991, Abys2010} have reported that incorporating 20\% Ni by weight ($\sim$31\% by mole) improves 
hardness, ductility and resistance to hydrogen embrittlement. This may indicate enhanced stability in that composition
range. Finally, our convex-hull diagram also agrees well with previously published data, for instance with the diagram 
generated using the Aflowlib library~\cite{Aflowlib}. 

% -----------------------------------------------------------------------------
\subsection{Vibrational free energy}
An accurate prediction of the phase stability of disordered alloys requires accounting for finite-temperature 
effects beyond simply considering the configurational entropy. In particular, the vibrational entropy can 
determine whether a phase is stable at finite temperatures and its calculation is thus 
fundamental~\cite{Fultz2010, Walle2002}. For example, the vibrational entropy stabilizes the L1\textsubscript{2} 
phase in Co-Al-W alloys at elevated temperatures \cite{Rhein2015}, while vibrational and configurational 
entropy together influence the phase stability across temperature in TiAl-based HEAs \cite{Hatzenbichler2023}.
The calculation of the vibrational contribution to the entropy is, however, numerical demanding. As such, several
approximations have been designed. For instance, a commonly used and lightweight method is rooted in the 
Debye-Gr\"uneisen approach, which is widely adopted for its computational efficiency 
\cite{Rogal2017,Hatzenbichler2023,Ma2015,Korman2016}. This, however, often does not capture important 
vibrational features in disordered alloys, a shortfall that can lead to inaccurate free-energy predictions 
\cite{Fultz2010}. More accurate vibrational free-energy calculations are typically based on the harmonic 
approximation, where the phonon spectra are computed from the dynamical matrix. Unfortunately, DFT 
phonon calculations for disordered alloys are computationally expensive due to the large supercells one 
has to consider and the many atomic configurations required for an appropriate sampling.

In this work, the vibrational free energy, $F^\mathrm{vib}$, is computed within the harmonic approximation 
by using the finite-displacement method as implemented in Phonopy (see Section \ref{sec:DFT}). Here, 
however, DFT is replaced by our JLP, which is employed to efficiently evaluate the atomic forces for a large 
number of configurations. The vibrational free energy for the $\sigma$ configuration of the alloy of lattice type 
$\alpha$ is written as 
\begin{equation}
F^\mathrm{vib}_{\alpha}(\sigma, T) = \frac{1}{2}\sum_{\textbf{q}\nu} \hbar \omega(\textbf{q}\nu) 
+k_\mathrm{B}T \sum_{\textbf{q}\nu} \ln[1-\mathrm{e}^{-\hbar\omega(\textbf{q}\nu)/k_\mathrm{B}T}],
    \label{eq:vib_free_energy}
\end{equation}
where $\omega(\textbf{q}\nu)$ is the phonon frequency for the mode with wave vector $\textbf{q}$ and band 
index $\nu$, while $k_\mathrm{B}$ is the Boltzmann constant. Since the free energy is specific of each
configuration $\sigma$ and temperature dependent, it is calculated for each structure as a function of $T$.
Phonon calculations are performed for all structures in the test set {\bf 1}, where the forces of the cells with
atoms displaced from their equilibrium position are evaluated using the JLP. To ensure no negative phonon 
frequencies and for consistency, the JLP-relaxed structure is taken as the equilibrium structure.
For comparison we also perform 42,067 DFT total-energy calculations, to obtain the phonon spectrum and 
the free energy of 673 different atomic arrangements. This set forms the DFT phonon reference data for test 
set {\bf 1}.
\begin{figure*}[htp]
    \centering
    \includegraphics[width=\linewidth]{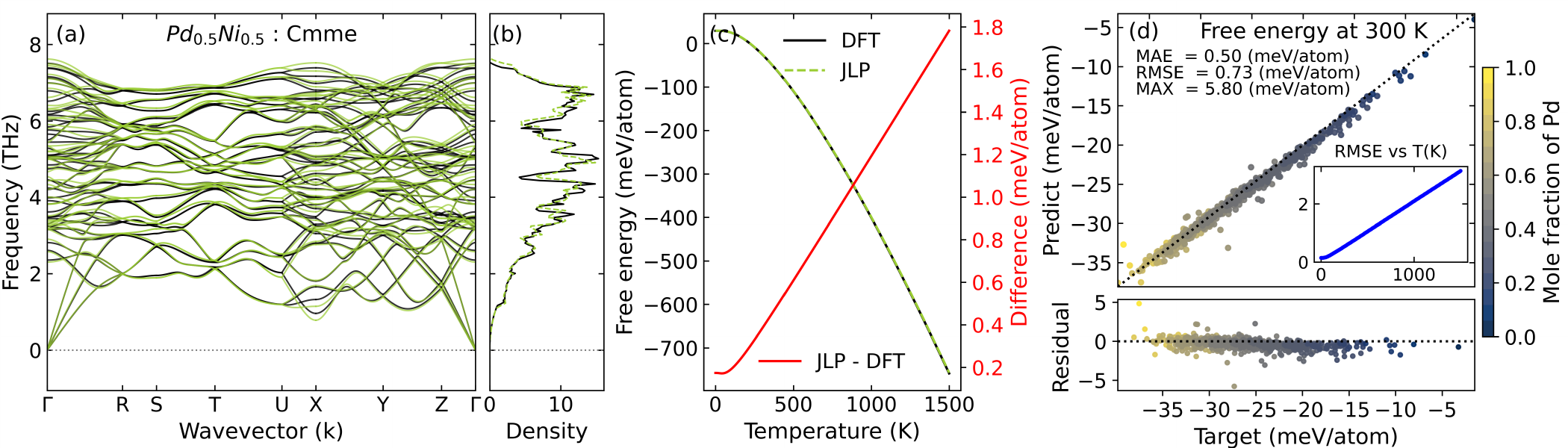}
    \caption{Summary of the phonon calculations performed. (a) Band structure, (b) density of states (DOS) and
    (c) vibrational free energy as a function of temperature of a selected NiPd structure with {\it Cmme} symmetry. 
    In all cases we show DFT and JLP results, and in panel (c) we also report the difference between the JLP
    free energy and that computed by DFT (red line and right-hand side scale). In panel (d) the parity plot of the 
    vibrational free energies at 300~K, comparing JLP and DFT results for test set {\bf 1}. In the inset we report the 
    RMSE at different temperatures.}
    \label{fig:compare_phonon}
\end{figure*}

As shown in Figs.~\ref{fig:compare_phonon}(a) and \ref{fig:compare_phonon}(b) for a representative example 
(with the remaining cases provided in the Supplementary Information), the phonon band structure and DOS 
computed by JLP agree well with the DFT reference. Small inaccuracies are largely compensated in the 
calculation of the free-energy, Fig.~\ref{fig:compare_phonon}(c), which for the example provided has a 
deviation from the DFT value of only 1.78~meV/atom at 1500~K. This corresponds to a relative error of 
0.25\%.
Similar results are observed across the entire test set {\bf 1}, as shown in Fig. \ref{fig:compare_phonon}(d). 
We find a RMSE of 0.64~meV/atom at 300~K, a value that increases linearly to 3.14~meV/atom at 1500~K 
(see the inset). This linear dependence is largely due to the linear scaling of the free energy with $T$ in 
Eq.~\ref{eq:vib_free_energy}, which also suggests that the error on the predicted vibrational entropy itself 
is constant at different temperature. Given the excellent performance of the JLP on test set {\bf 1}, we extend 
the phonon calculations to the entire dataset, by spanning all mixtures accessible with 2$\times$2$\times$3 
supercells. This involves performing 14,294,592 force evaluations, a task completely out of reach of DFT alone.
The computed free energy are then used to construct the temperature-dependent Gibbs free energy of formation 
and therefore establish the thermal stability of a particular phase or solid-state solution.

% -----------------------------------------------------------------------------
\subsection{Thermodynamics phase stability}\label{sec:phase_stability}
The phase stability of a multicomponent alloy of lattice type $\alpha$ with a given mixing composition 
$\{x_i^\alpha\}$ at pressure $P$ and temperature $T$ is determined by the associated Gibbs free energy. 
By using the notation introduced in reference~\cite{Ikeda2019}, the Gibbs free energy is obtained from the 
Helmholtz free energy, $F_\alpha$, as
\begin{equation}
    G_\alpha(\{x_i^\alpha\}, P, T) = F_\alpha(\{x_i^\alpha\}, V, T) + PV\:.
    \label{eq:Gibbs}
\end{equation}
Given the set of DFT energies, $E_\alpha$, for all possible configurations $\sigma$ of a multicomponent alloy with 
mixing composition $\{x_i^\alpha\}$, the canonical partition function writes
\begin{equation}
    Z_\alpha(\{x_i^\alpha\}, T) = \sum_{ \sigma \in \{x_i^\alpha\}} \exp\left( -\frac{E_\alpha(\sigma)}{k_\mathrm{B}T} \right)\:.
    \label{eq:partition_func_wihout_vib}
\end{equation}
In this work, we account only for the configurational and vibrational contributions to the free energy, as these 
are generally the dominant thermodynamic drivers for the phase stability in alloy systems~\cite{Tolborg2022}. 
Where necessary, suitable approximations for the electronic and magnetic components of the free energy can 
be utilized, as demonstrated, for instance, in reference~\cite{Ikeda2019}.

The vibrational degrees of freedom are adiabatically decoupled from the configurational ones, since there is a 
large difference between the typical time scale of the atomic vibration and that of atomic diffusion. 
In this case, the partition function can be decoupled into a configurational and vibrational contribution \cite{Fontaine1994}, 
an approximation known as the partition-function coarse graining \cite{Walle2002},
\begin{equation}
\begin{aligned}
    Z_\alpha(\{x_i^\alpha\}, T) &= \sum_{ \sigma \in \{x_i^\alpha\}} \exp\left( -\frac{E_\alpha(\sigma)}{k_\mathrm{B}T} \right)  Z^\mathrm{vib}(\sigma, T) \\ & = \sum_{ \sigma \in \{x_i^\alpha\}}  \exp\left(-\frac{E_\alpha(\sigma) + F_\alpha^\mathrm{vib}(\sigma, T)}{k_\mathrm{B}T} \right)\:, 
\end{aligned}
\label{eq:partition_func}
\end{equation}
where $E_\alpha(\sigma)$ is the DFT (or JLP) energy of the $\sigma$ configuration at equilibrium, and 
$F_\alpha^\mathrm{vib}(\sigma, T)$ is its associated vibrational free energy as computed with 
Eq.~(\ref{eq:vib_free_energy}). Here,
\begin{eqnarray}
% Z(\sigma)= \exp\left( -\frac{E_\alpha(\sigma)}{k_\mathrm{B}T} \right)\:, \\
Z^\mathrm{vib}(\sigma, T)=\sum_{\omega}\exp\left(- \frac{E^\mathrm{vib}_{\alpha, \omega}(\sigma)}{k_\mathrm{B}T}\right)\:,
\end{eqnarray}
is the vibrational contribution to the partition function, with $E^\mathrm{vib}_{\alpha, \omega}(\sigma)$ being the 
energy of the $\omega$ vibrational mode of the $\sigma$ configuration. Essentially, in Eq.~(\ref{eq:partition_func}), 
the Boltzmann weight associated to the $\sigma$ configuration is modified by the addition of the vibrational free 
energy, $F_\alpha^\mathrm{vib}(\sigma, T)$, to the original configurational energy, $E_\alpha(\sigma)$.
Finally, the Helmholtz free energy can be directly obtained as
\begin{equation}\label{eq:Falpha}
    F_\alpha(\{x_i^\alpha\}, T) = -k_\mathrm{B} T \ln{Z_\alpha(\{x_i^\alpha\},T)}\:.
\end{equation}
At this stage, the volume dependence is neglected, as the 0~K minimum‑energy volume is employed for both 
the static‑energy and phonon calculations at all temperatures. The method used to include an explicit volume 
dependence, corresponding to the quasi‑harmonic approximation, will be discussed in section~\ref{sec:QHA}.
Since in solid-state systems the difference in volume is typically small (for the same lattice type), the $PV$ term 
in the Gibbs free energy, Eq.~(\ref{eq:Gibbs}), can be neglected,so that $G_\alpha$ can be approximated by 
the Helmholtz free energy, Eq.~(\ref{eq:Falpha}). 
Notably, our computational strategy does not rely on approximating the configurational entropy using either 
the ideal-mixing expression $S_\alpha(\{x_i^\alpha\})=-k_B\sum_i^Nx_i^\alpha \ln x_i^\alpha$
or entropy estimates derived from the configurational density of states~\cite{Zhang2024}. Instead, by explicitly 
enumerating all the symmetrically distinct configurations, we obtain the exact configurational entropy for that 
finite system.

\begin{figure}[htp]
    \centering
    \includegraphics[width=\linewidth]{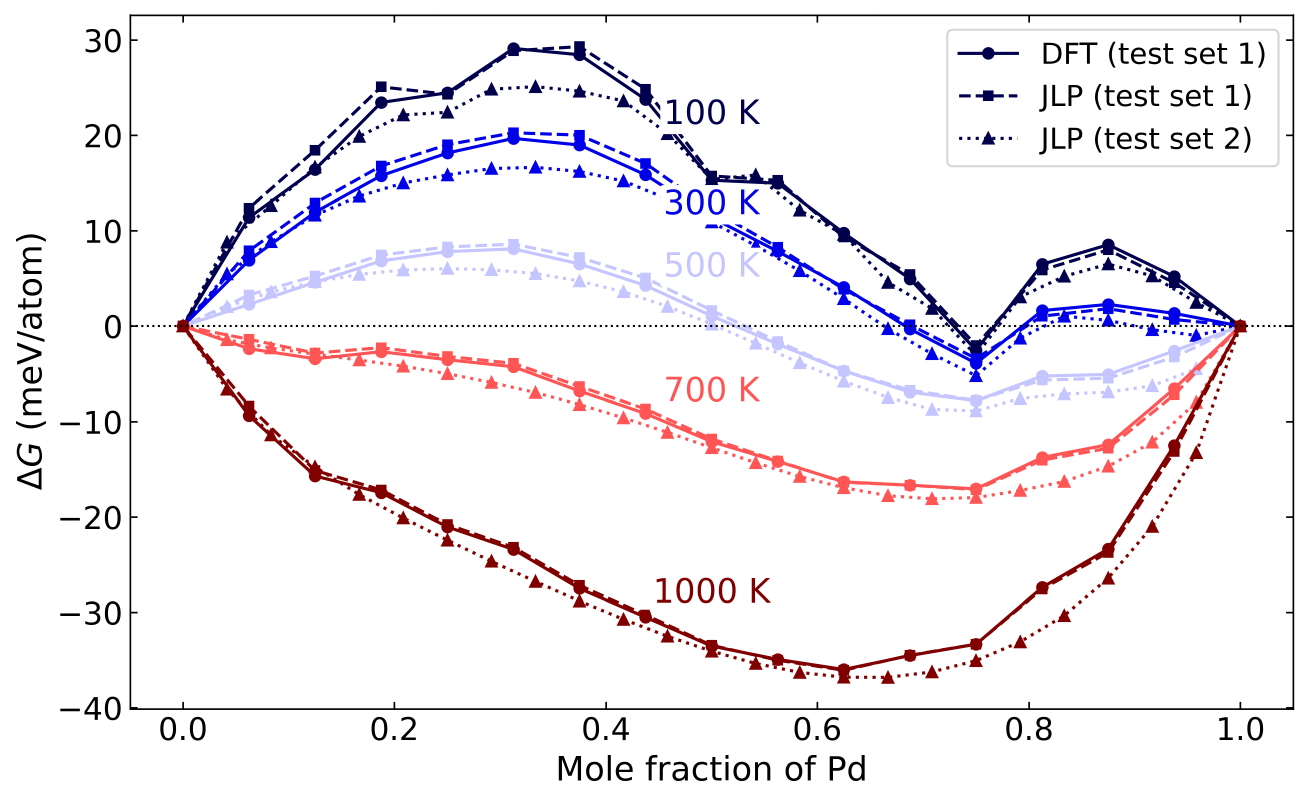}
    \caption{Temperature-dependent Gibbs free energy of formation. The Gibbs free energy of formation of NiPd 
    alloys is computed from the JLP across the entire composition range. For compositions compatible with a 
    16-atom supercell (test set {\bf 1}) results are also presented at the DFT level. Note that alloys become
    thermodynamical stable at all composition for temperatures greater than 700~K.
    }
    \label{fig:compare_gibbs}
\end{figure}

We analyse all possible configurations obtainable with either a 16- or a 24-atom supercell, which amount at
65,536 and 16,777,216, respectively. Symmetry identification allows us to consider only symmetry‑unique 
structures, with the contribution of each accounted for by its multiplicity (as described in section~\ref{sec:test_structures}). 
This reduces the total number of structures that must be evaluated to 672 and 99,268, respectively.
These structures cover a fine range of possible stoichiometry across the binary phase diagram.
The Gibbs free energy of formation, $\Delta G$, is then computed with the JLP with respect to the 
boundary unary phases, as shown in Fig.~\ref{fig:compare_gibbs}. Furthermore, for the compositions
accessible with the 16-atom cells, the same calculation is repeated at the DFT level.

From the figure, it emerges that {\it fcc} Ni-Pd alloys are thermodynamically stable across the entire composition 
range for temperatures above a critical temperature found in between 500~K and 700~K. Since the lowest
melting point is found at 1,237~$^\circ$C for an alloy with 45\% Pd atomic fraction~\cite{Nash1984,Fraenkel1927},
we conclude that the alloys are thermodynamically stable when quenched from the liquid phase,
to become metastable at low temperature. This confirms the miscibility of Ni and Pd across the 
entire composition range. Interestingly, Pd-rich alloys retain thermodynamical stability at temperatures
lower than those needed by the Ni-rich side of the composition range. 

In general, our JLP-computed Gibbs energies agree well with those obtained by DFT across all mixing 
compositions and temperatures. Note that we compare directly only for test set {\bf 1}, namely for 16-atom
cells. When looking at the JLP results obtained with the 24-atom cells, we note that some small deviations
emerge, and in particular $\Delta G$ remains systematically lower than that obtained from the 16-atom
cells. This should not be surprising. The configuration entropy, in fact, is proportional to the number of
available configurations~\cite{He2016}, $\Omega$, which in turn is determined by the number of atoms, $N$, in the 
supercell considered. For example, there are $\Omega = \binom{N}{N/2}$ possible configurations 
for the half-mixing binary alloys for a cell of $N$ atoms, and the associated maximum configurational entropy 
(in the high-temperature limit, where all configurations are equally probable) is $S = \frac{k_B}{N}\ln\Omega$.
The entropy then reaches the ideal-mixing value ($S = k_B \sum_i x_i \ln x_i$) for $N\rightarrow\infty$, as 
further discussed in section~\ref{sec:ideal_entropy}.
As a consequence, the 24-atom supercells provide a more accurate estimate of the configurational entropy, 
and in general a larger $TS$ contribution to the Gibbs free energy than the 16-atom cells. For this reason
$\Delta G$ remains systematically lower. However, these deviations are minor and we can conclude that
our results already provide an excellent description of the finite-temperature phase stability of the Ni-Pd
system.

% -----------------------------------------------------------------------------
\subsection{Effect of the vibrational Gibbs energy}
According to statistical mechanics, the partition function, $Z$, is the central quantity that links the 
microscopic states of a system to its macroscopic thermodynamic properties. This is defined through 
the cDOS, which thus fully determines the thermodynamics of a system~\cite{Sutton2020}.
Once $Z$ is known, all equilibrium quantities, such as the Gibbs free energy, the internal energy, 
the entropy and the heat capacity, can be obtained from its derivatives with respect to temperature 
and other variables. In contrast to sampling-based approaches, our method compute $Z$ without 
requiring independent simulations for each temperature and composition. In fact, by enumerating all 
unique configurations, we obtain the exact cDOS and partition function (for given finite supercell size), 
allowing direct calculation of the thermodynamic properties across a range of conditions.

In order to understand how the various degrees of freedom determine the thermodynamics properties 
across the Ni-Pd phase diagram, in this section, we evaluate the configurational entropy and the heat 
capacity by either considering or neglecting the vibrational contributions to the partition function. This 
is performed for the 24-atom supercells (test set {\bf 2}). We first consider the partition function in the 
absence of lattice vibrations, so that $Z$ is determined solely by the static energies (configurational
entropy only), as given in Eq.~(\ref{eq:partition_func_wihout_vib}). This can be compared with the case 
where the free energy comprises both static and vibrational terms, as defined in Eq.~(\ref{eq:partition_func}).
\begin{figure}[htp]
    \centering
    \includegraphics[width=0.85\linewidth]{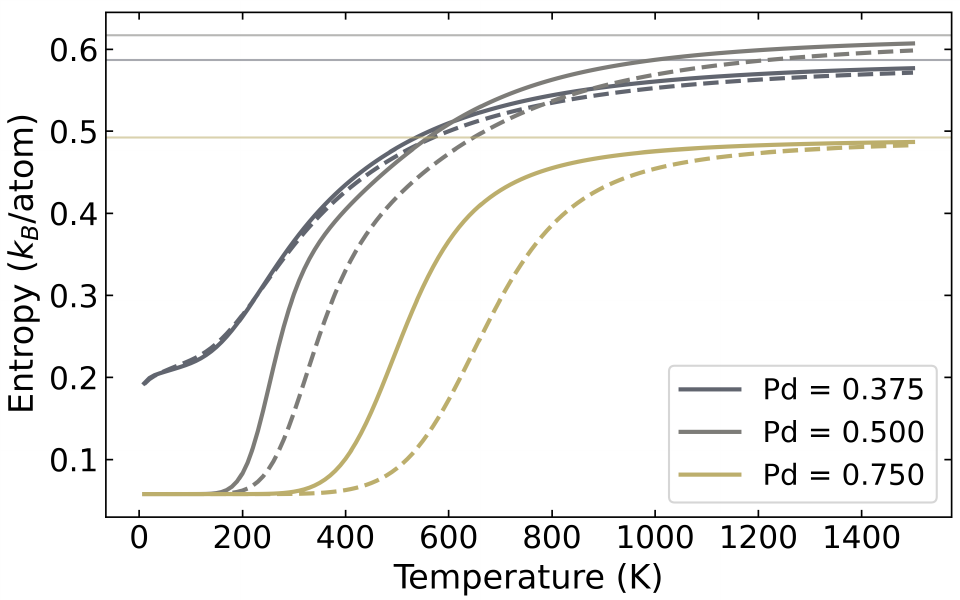}
    \caption{Configurational entropy obtained with (solid lines) and without (dashed lines) the inclusion of 
    the vibrational contributions to the free energy. Results for different mixing compositions are shown by 
    different colors. The dotted lines indicate the maximum configurational entropy, ideal mixing configuration
    entropy, corresponding to the fully random limit for the finite supercell size used in the calculations.}
    \label{fig:entropy}
\end{figure}

Using the entropy relation, 
\begin{equation}
    S(\{x_i\}, T) =  \frac{U(\{x_i\},T)-F(\{x_i\},T)}{T}\:
    \label{eq:entropy}
\end{equation}
where $U$ denotes the internal energy, the configurational entropy obtained with and without 
vibrational contributions is compared in Fig.~\ref{fig:entropy}. Our results show that lattice 
vibrations significantly impact the system entropy, up to about 0.1~$k_{\mathrm{B}}$, in the 
case of Pd-rich alloys, while the effect is much smaller for the Ni-rich case. This is consistent 
with the experimentally reported Debye temperature, $\Theta_D$, which decreases with increasing 
the Pd content~\cite{Mackliet1963}. The lower $\Theta_D$ in Pd-rich alloys, driven by the larger 
atomic mass and softer bond stiffness of Pd relative to Ni, results in the more pronounced vibrational 
entropy contribution observed here.
The vibrational contribution to the entropy is maximized at intermediate 
temperatures, while it plays a small role both in the low- and high-temperature limit. At intermediate
temperature the local chemical environments fluctuate strongly, so that the excess configurational 
entropy arises from vibrational free‑energy differences among configurations. Notably, the total
entropy reaches up the ideal mixing limit at progressively higher temperature as the Pd fraction
in the alloy is increased. Since, as observed in Fig.~\ref{fig:compare_gibbs}, Ni-Pd alloys become 
thermodynamically stable at lower temperature first in the Pd-rich part of the phase diagram, we 
conclude that the stability is driven by entropy at the Ni side, and by enthalpy at the Pd one. 
At high temperatures, where the alloy approaches the ideal mixing limit, the vibrational free energies 
become nearly configuration‑independent and therefore do not contribute to configurational entropy 
differences. It is important to note the total $S$ reaches its maximum value of $\approx 0.62\ k_{\mathrm{B}}/\text{atom}$ 
at half mixing (as shown by the dotted-lines in Fig.~\ref{fig:entropy}). This is just slightly below the 
ideal-mixing limit for equiatomic binaries, $0.693~k_{\mathrm{B}}/\text{atom}$, due to the finite size
of our simulation cell (see Appendix~\ref{sec:ideal_entropy} for further details).
\begin{figure}[htp]
    \centering
    \includegraphics[width=0.85\linewidth]{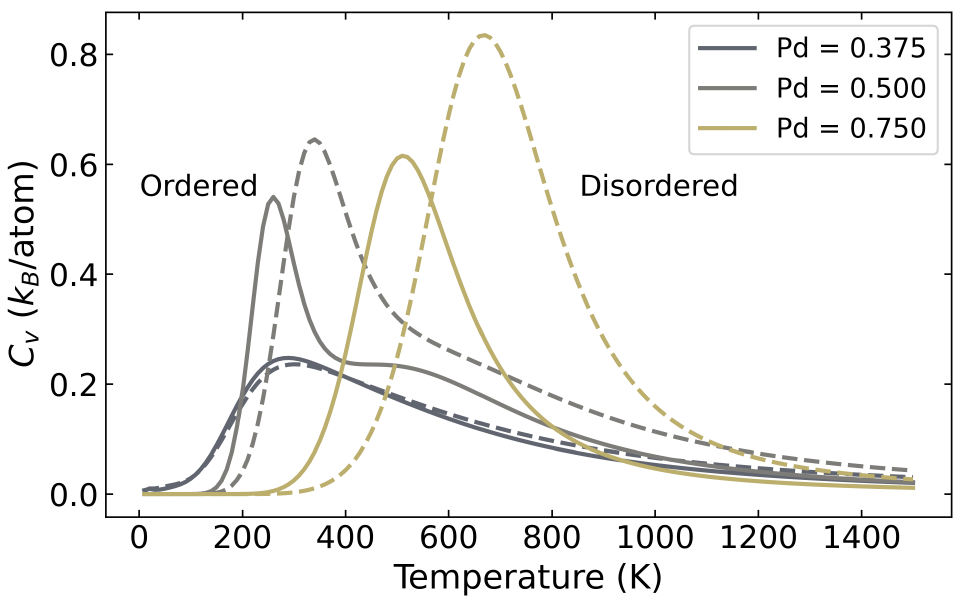}
    \caption{Heat capacity obtained with (solid lines) and without (dashed lines) including the vibrational 
    contribution to the free-energy. Results for different mixing compositions are shown with different colors. 
    The peaks in the heat-capacity are found at the temperature where the configurational energy fluctuations 
    are maximized. Thus $C_V(\{x_i\},T)$ provides an estimate of the order--disorder transition temperatures.
    }
    \label{fig:heat_capacity}
\end{figure}

Next we examine the heat capacity, $C_V(\{x_i\},T)$, a quantity that provides insight into the temperature ranges 
over which the system exhibits long-range order, short-range order, or a disordered state. This is computed using 
the definition 
\begin{equation}
C_V(\{x_i\},T)
= \frac{\partial U(\{x_i\},T)}{\partial T}\:.
\label{eq:heat_capacity}
\end{equation}
The heat capacity as a function of temperature typically shows a peak corresponding to the temperature 
at which the energy fluctuations are maximized. This peak effectively marks the boundary between ordered
and disordered structures. 
Results obtained respectively at 37.5\%, 50\% and 75\% Pd content are shown in Fig.~\ref{fig:heat_capacity}.
Let us discuss the 50\% and 75\% cases first. For both concentrations, the lowest-energy configuration adopts 
a LP--L1$_0$ structure (see Fig.~\ref{fig:convex_hull}), which is separated from the next competing phases 
by a relatively large energy gap ($\approx$10~meV/atom for 75\% Pd). This implies that substantial thermal 
energy is required to thermally populate excited configurations. As a consequence, for these concentrations
the heat capacity remains nearly constant at low temperature (up to approximately 200~K for 50\% Pd
concentration and 400~K for 75\%). Then, above such threshold temperatures excited configurations become 
thermally accessible, resulting into large energy fluctuations and the corresponding peak in the heat capacity.
In contrast, for 37.5\% Pd content the configurational energies are more continuously distributed, allowing 
excited configurations to be populated over a broader temperature range. As a result, the heat capacity varies 
more smoothly with temperature and does not exhibit the sharp peaks observed for the 50\% and 75\%
cases.

The inclusion of vibrations (solid lines in Fig.~\ref{fig:heat_capacity}), shifts the $C_V(\{x_i\},T)$ peaks to lower 
temperature: from approximately 700~K to 500~K for 75\% Pd content and from about 350~K to 250~K for 50\%. 
This means that the order--disorder transition shifts to lower temperature due to the thermal vibrations.
Such behaviour is expected, since typical vibrational entropy differences in binary alloys are reported to be
of the order of 0.1--0.2~$k_{\mathrm{B}}$ per atom, values that can lead to corrections of up to approximately 
30\% in predicted order--disorder transition temperatures~\cite{Walle2002}.

As emphasized in~\cite{Ikeda2019}, the thermodynamic description of alloys becomes increasingly complex 
near the order--disorder transition temperature, where multiple competing configurations and pronounced 
chemical short-range order can contribute simultaneously to the free energy. 
In this regime, single-configuration approximations, such as SQS, may fail to 
fully capture the configurational diversity of the system. In these 
approximations, vibrational contributions are evaluated for a single 
representative atomic configuration and the configurational degeneracy is 
incorporated through the configurational entropy approximation 
$F = E - TS_{\text{conf}} - F_{\text{vib}}$.
In contrast, the efficiency and accuracy 
of the JLP enable the explicit sampling of a configurational ensemble, providing a unified description of long-range 
ordered, short-range disordered, and disordered states, and allowing configuration-dependent energetic and 
vibrational effects to be captured reliably across the entire temperature range.
% -----------------------------------------------------------------------------

\subsection{Mechanical properties}\label{sec:mechanical}
\begin{figure}[htp]
    \centering
    \includegraphics[width=1\linewidth]{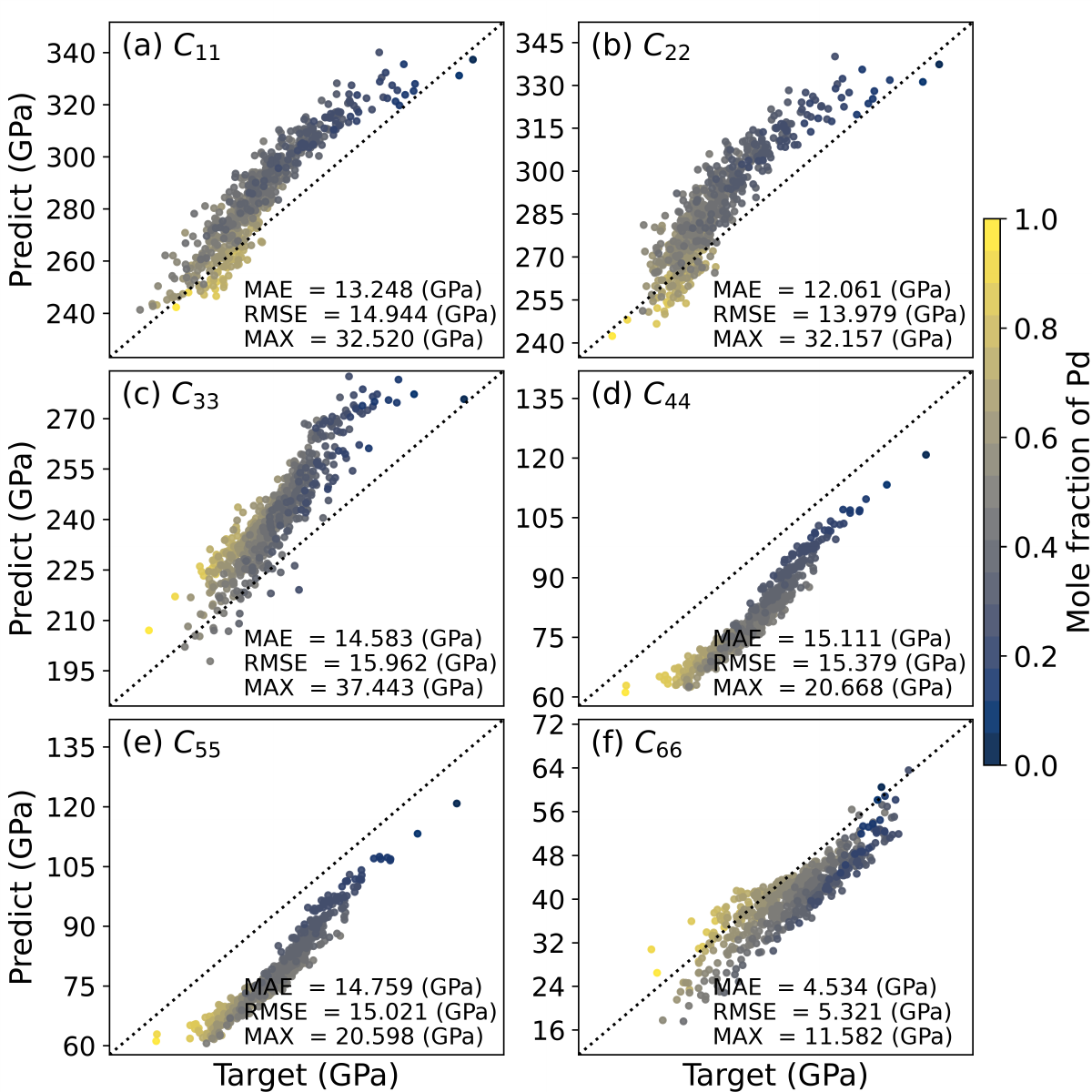}
    \caption{Parity plots for the diagonal components of the elastic matrix computed at the DFT ($x$-axis) and 
    JLP ($y$-axis) level for test set {\bf 1}. We report (a) C$_{11}$, (b) C$_{22}$, (c) C$_{33}$, (d) C$_{44}$, 
    (e) C$_{55}$ and (f) C$_{66}$. On each graph we report the MAE, RMSE and the maximum error (MAX).}
    \label{fig:elastic_const}
\end{figure}

Mechanical properties are critical for the practical application of alloys, especially HEAs, where a balance 
of strength, ductility and thermodynamics stability is often required \cite{Lach2025}. The elastic tensor describes 
how a material responds to external forces within the elastic limit, and it contains information about various 
mechanical responses such as the Young modulus, bulk modulus, and Poisson ratio. These are difficult to compute
across composition and temperature, since large configuration averages are needed, a task suitable for the
JLP. By definition, the stress-strain relation in the linear regime is described by a $6\times6$ symmetric elastic 
matrix $\bm{C}$
\begin{equation}
    \bm{\sigma} = \bm{C} \bm{\varepsilon}\:,
\end{equation}
where $\bm{\sigma}$ contains the 6 independent strain tensor components and their corresponding strain 
deformations are $\bm{\varepsilon}$. The stress tensor is then given by
\begin{equation}
    \sigma_{\mu \nu} = \frac{1}{V} \frac{\partial E}{\partial \varepsilon_{\mu\nu}},
\end{equation}
where $\mu$ and $\nu$ label the independent Cartesian directions $x$, $y$ and $z$. Following the approach 
developed by Materials Project \cite{deJong2015} as implemented in {\sc Pymatgen} \cite{pymatgen}, we compute 
the elastic tensor by fitting stress-strain data from both normal and shear deformations in the range of $-1\%$, $-0.5\%$, 
$+0.5\%$, and $+1\%$. 
DFT calculations are performed for all 16,152 deformed structures, including full atomic relaxation, and the elastic 
tensors are calculated as reference values for test set {\bf 1}. The same workflow is then carried out using JLP, starting 
from the JLP-relaxed structures. The resulting diagonal components of the elastic matrix are compared in 
Fig.~\ref{fig:elastic_const}. 

For $C_{11}$, $C_{22}$, and $C_{33}$, which correspond to normal strain, the JLP RMSEs are within $\sim$10\% of 
the DFT values. In contrast, the remaining components, which correspond to shear strains, present considerably larger
errors. These arise from the lack of shear-strained structures in the training data and from the inclusion of stress values 
far beyond the elastic regime used for the elastic tensor calculations (see Fig.~\ref{fig:stress_dist}). The accuracy can 
then be improved by incorporating shear-strained unit cells and by reducing the applied strain used in the training set, 
as we will discuss in Appendix~\ref{sec:improve_elastic}.

The thermodynamic average over all configurations $\sigma$ of the elastic constant $C_{ij}$ of disordered alloys at 
different mixing composition $\{x_i\}$ and temperatures $T$ can be evaluated as
\begin{equation}
    \langle C_{ij}(\{x_i\}, T) \rangle = \sum_{ \sigma \in \{x_i\}} p(\sigma, T) C_{ij}(\sigma,),
    \label{eq:average_elastic}
\end{equation}
where $p(\sigma, T)$ is the Boltzmann probability of configuration $\sigma$, determined from the partition function of 
Eq. \ref{eq:partition_func}. 
\begin{figure}
    \centering
    \includegraphics[width=\linewidth]{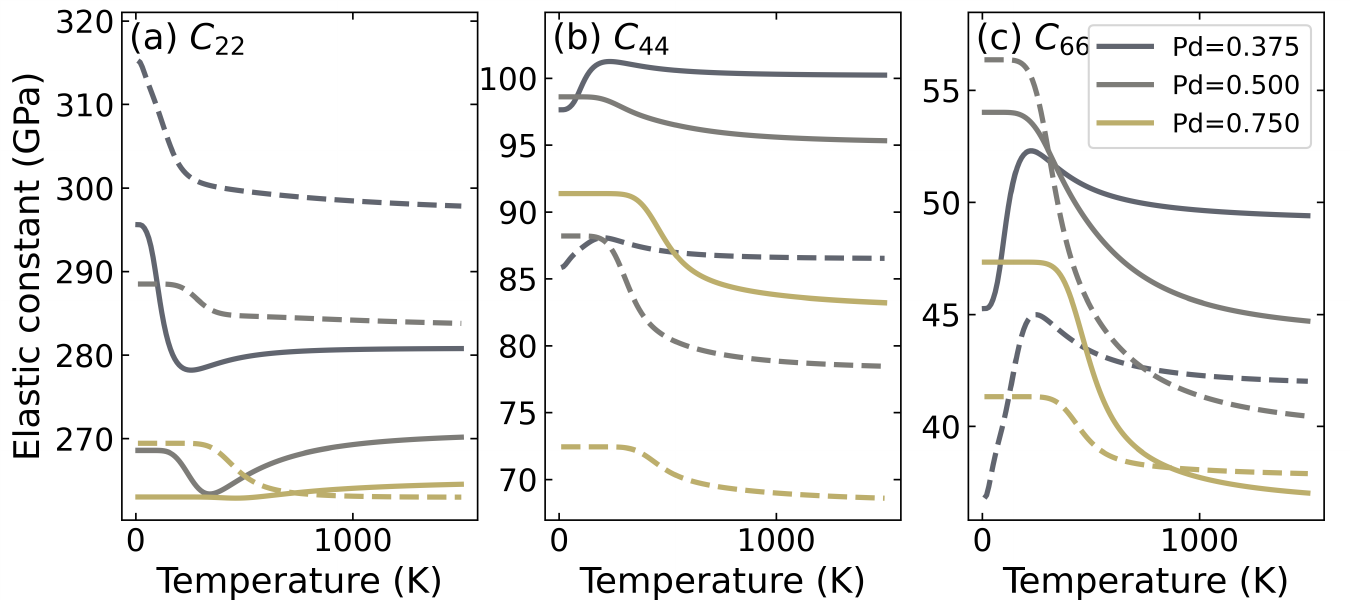}
    \caption{Thermodynamically averaged elastic constants $\langle C \rangle$ as a function of 
    temperature for the Pd--Ni system at Pd compositions of 37.5\%, 50\%, and 75\%. Panels (a), 
    (b), and (c) show $\langle C_{22} \rangle$, $\langle C_{44} \rangle$, and $\langle C_{66} \rangle$, 
    respectively. The averages are obtained via canonical ensemble (Boltzmann) averaging over 
    configurational energies.
    Solid lines are for the DFT results, whereas dashed lines represent the JLP ones.
    }
    \label{fig:average_elastic}
\end{figure}
The resulting thermodynamic averages are shown in Fig.~\ref{fig:average_elastic} for Ni--Pd alloys with Pd 
fractions of 37.5\%, 50\%, and 75\%. We observe pronounced variations in $\langle C_{ij} \rangle$ near the
configurational transition temperatures identified from the heat capacity (see Fig.~\ref{fig:heat_capacity}).
For most compositions and elastic components, the trends predicted by the JLP model follow those obtained 
directly from DFT, with similar patterns of underestimation/overestimation as seen in Fig.~\ref{fig:elastic_const}.
Thus, the error that we observe in the thermodynamical average of the elastic constants simply reflect that 
found for single configurations.

In the current  approach, the temperature dependence of the averaged elastic constants arises solely from 
the configurational entropy via Boltzmann weighting of the atomic configurations. The elastic constants, 
$C_{ij}(\sigma)$, entering Eq.~\ref{eq:average_elastic} are all evaluated at the 0~K equilibrium volume.
This means that explicit vibrational contributions and thermal expansion effects are not included at this 
point. Typically, the elastic constants are expected to decrease with increasing temperature due to 
volumetric elastic softening caused by the thermal expansion. This effect clearly cannot be captured when 
the elastic constants are evaluated at a fixed 0~K volume. The temperature dependence of the elastic 
constants can, in general, be obtained within the quasistatic approximation using thermal expansion 
coefficients and thermodynamic quantities from quasi-harmonic calculations, followed by an isothermal--adiabatic 
conversion, as implemented for example in \textsc{Elastemp}~\cite{Balasubramanian2023}.
To this end, quasi-harmonic approximation calculations using the JLP are presented in 
Section~\ref{sec:QHA}.

Certainly, the results presented in Fig.~\ref{fig:average_elastic} are qualitative satisfactory, but quantitatively
noisy with respect to the DFT reference. It is, however, important to emphasize that the primary objective of 
this work is to obtain accurate finite-temperature phase stability and reasonable estimates of mechanical 
properties without retraining the model. To this end, the training data are intentionally sampled only around 
equilibrium configurations, with an emphasis on minimizing the amount of DFT data required for training the
model. In particular, strained unit cells are used in order to reduce computational cost and to ensure that the 
effort to generate the training data does not become prohibitively expensive when extending the approach 
to multicomponent alloy systems.
A larger training dataset will very likely lead to significant quantitative improvements. For example, including 
additional normal and shear strains for each unique configuration, or generating more atomic perturbations 
and strain states for each configuration, would enhance the model's ability to generalize across a wider range 
of volumes and deformations. Such data-generation strategies have been demonstrated to improve transferability 
in recent work, for instance in reference.~\cite{Poul2025}.
It should also be noted that elastic constants are intrinsically more challenging to predict than energies, as they 
depend on the curvature of the potential energy surface rather than its absolute energy value. Moreover, in this 
work, elastic constants are evaluated exclusively from equilibrium structures optimized by JLP and are obtained 
by fitting the stress--strain relationship. As a consequence, even small residual errors in the predicted equilibrium 
positions or volumes can be amplified through the stress calculation, leading to noticeably larger deviations in the 
resulting elastic constants.

% -----------------------------------------------------------------------------
\subsection{Quasi-harmonic approximation}\label{sec:QHA}
\begin{figure}[htp]
    \centering
    \includegraphics[width=\linewidth]{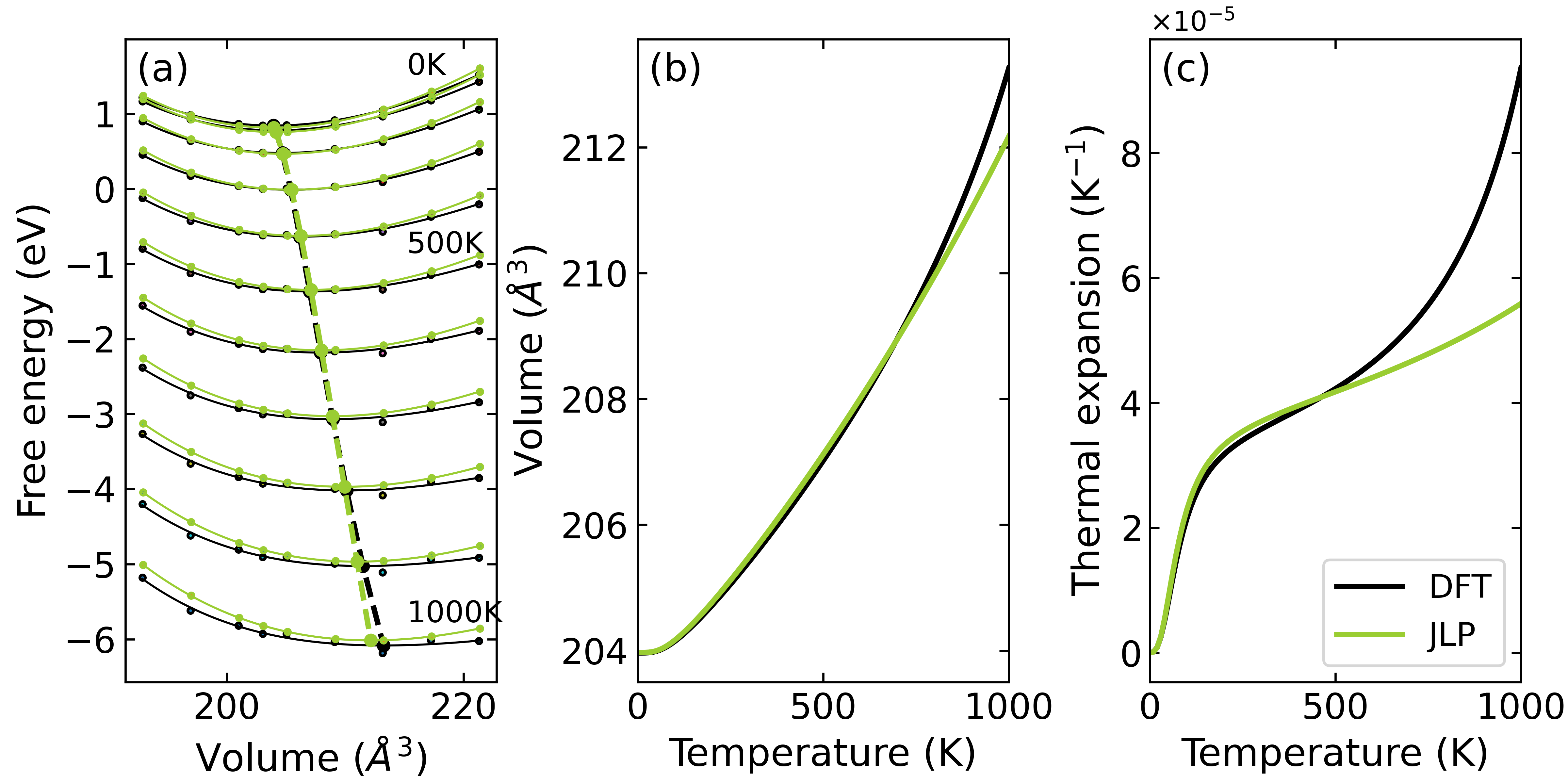}
    \caption{Quasi-harmonic approximation results at zero external pressure for the lowest-energy 
    configuration at 37.5\% Pd composition. (a) Free energy as a function of temperature going from 
    0 to 1000~K, with the dashed line indicating the equilibrium volumes. (b) Temperature dependence 
    of the equilibrium volume. (c) Thermal expansion coefficient as a function of temperature. The DFT 
    results are depicted in black, while the JLP results are in green.}
\label{fig:QHA}
\end{figure}
The quasi-harmonic approximation (QHA) is an extension of the harmonic model that 
considers volume‑dependent phonon frequencies in order to capture the thermal expansion of 
crystalline solids. Within the QHA, finite-temperature thermodynamic properties are obtained by
minimizing the Helmholtz free energy with respect to the volume, where the free energy includes 
both static and vibrational contributions. For a given atomic configuration $\sigma$, the QHA free 
energy at temperature $T$ and pressure $p$ is defined as
\begin{equation}
F^{\mathrm{QHA}}(\sigma, T, p) =
\min_V \left[ E(\sigma, V) + F^{\mathrm{vib}}(\sigma, T; V) + pV \right],
\end{equation}
where the volume minimization returns to the equilibrium volume at temperature $T$. Although the 
QHA provides a more accurate description of finite-temperature effects than the harmonic approximation, 
it is computationally demanding. For each atomic configuration, QHA typically requires phonon calculations 
at multiple volumes (often 5--10), making it prohibitively expensive for first-principles DFT and large configurational 
datasets.

Using the \textsc{phonopy-qha} package~\cite{Togo2010}, we performed QHA calculations for all structures 
in test set~\textbf{1} at zero pressure using JLP, with a limited number of representative configurations also 
evaluated at DFT level for comparison. For clarity, we focus the discussion on the lowest-energy configuration 
at 37.5\% Pd composition as a representative case; results for all remaining configurations are provided in the 
Supplementary Information.
As shown in Fig.~\ref{fig:QHA}(a), the QHA free energies predicted by JLP agree well with the DFT results 
at low to intermediate temperatures. At higher temperatures, the deviation gradually increases, following a 
trend similar to that observed for the phonon free energies in Fig.~\ref{fig:compare_phonon}(c). The larger 
deviations observed for volumes far from equilibrium are also expected, since the JLP training dataset contains 
limited information on large volumetric expansions, aside from the strained unit-cell configurations. The temperature 
dependence of the equilibrium volume and thermal expansion coefficient, shown in Fig.~\ref{fig:QHA}(b) and (c), 
is smooth and physically reasonable up to intermediate temperatures. 

We than conclude that overall, the JLP provides a reasonable and computationally efficient
approximation to quasi-harmonic thermodynamics up to intermediate temperatures. Achieving higher 
accuracy at elevated temperatures will likely require augmenting the training set with configurations 
sampling a broader range of volumetric strains, together with corresponding displaced structures at each
volume to ensure accurate phonon properties across the QHA volume range.

% -----------------------------------------------------------------------------
\subsection{Discussion: extension to multiple atomic species}\label{sec:extension}
While the model presented here is constructed for two atomic species, it can be straightforwardly extended 
to additional species by incorporating the appropriate 2B and 3B clusters into the descriptor set. To discuss 
the computational scaling of this effort, one has to bare in mind that the number of model parameters increases 
linearly with the number of clusters, so that the total number of clusters determines the size of the model. Assuming 
to truncate the cluster expansion to 3B, the total number of clusters is given by $\frac{k^3}{2} + k^2 + \frac{k}{2}$, 
where $k$ is the number of unique atomic species. For example, the possible 2B clusters for three atomic species 
$A$, $B$ and $C$ are $A$-$A$, $B$-$B$, $C$-$C$, $A$-$B$, $A$-$C$ and $B$-$C$. In fact, the total number of 
2B terms is given by $k + \binom{k}{2} = k+ \frac{k(k-1)}{2}$. Similarly, the 3B clusters are determined by selecting 
a center atom type and by pairing it with the possible 2B clusters, so to obtain thus $k^2 + \frac{k^2(k-1)}{2} $ terms.
In short, extending a model from binary to ternary and quaternary alloys would theoretically require 2.67 and 5.56 
times more model parameters, respectively. 

\begin{figure}[ht]
    \centering
    \includegraphics[width=\linewidth]{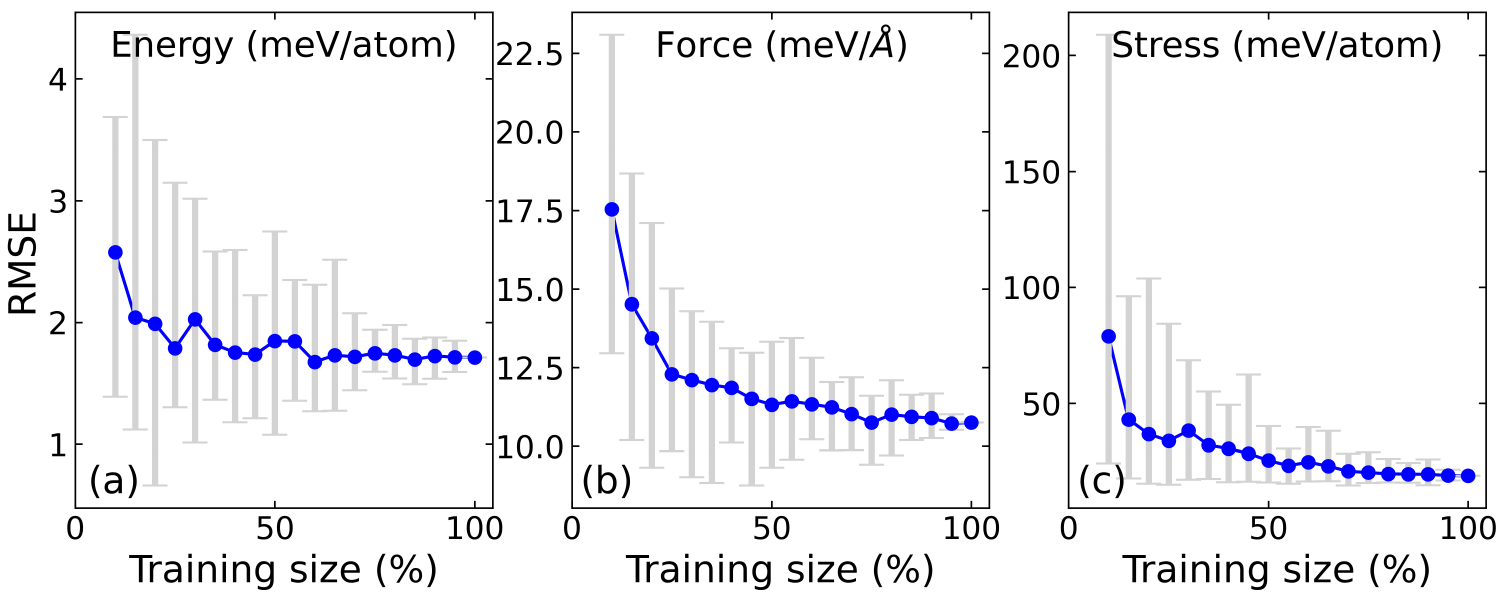}
    \caption{Convergence plot for energy (a), forces components (b) and stress (c) computed for a model constructed over 
    training sets of different size (from a total of 1059 training configuration). The error is evaluated over the test set {\bf 1}. 
    For each training size the errors are averaged over 20 random subsets, with error bars indicating the minimum and maximum 
    errors.}
    \label{fig:train_convergence}
\end{figure}

For the Ni-Pd binary alloys, we have demonstrated that a compact model with only 873 parameters, can be 
constructed to generalize well to larger supercells. Remarkably, the time required for the 14,294,592 force 
calculations needed for test set {\bf 2} is only 12 hours ($\approx$ 1.19 million calculations per hour), on a 
node equipped with two Intel Xeon E5-2699 v4 CPUs (44 cores total, 2.2 GHz) and 256 GB of RAM, running 
a 64-bit Linux system. In addition, all calculations can be performed independently once all the unique configurations 
are determined, allowing for efficient parallelization.  Regarding the training data, an energy accuracy of approximately 
2.58~meV/atom can be obtained with just 10\% of the training set (52 structures), while force and stress predictions 
converge at around 40\% of the training set, as shown in Fig. \ref{fig:train_convergence}. 
This suggests that the model requires only a small amount of training data, and that the dataset used here is not 
optimal, since a similar accuracy could likely be achieved with a smaller but more representative training set. 
Advanced techniques such as active learning \cite{MTP1,Jinnouchi2019,Vandermause2020,Sharma2024} can be 
employed to optimally select training structures for DFT labelling, thereby significantly reducing computational cost, 
especially when extending to multiple atomic species. Furthermore, regularization methods can be applied to reduce
model complexity and improve transferability. These aspects will be addressed in a future work.

%%%%%%%%%%%%%%%%%%%%%%%%%%%%%%%%%%%%%%%%%%%%%%%%%%%%%%%%%%%%%%%%
%%%%%%%%%%%%%%%%%%%%%%%%%%%%%%%%%%%%%%%%%%%%%%%%%%%%%%%%%%%%%%%%

\section{Conclusion}\label{sec:Conclusion}

We have demonstrated a complete workflow for the construction of a machine-learning interatomic potential, 
the JLP, for the calculation of the phase stability and mechanical properties of disordered alloys. In particular,
we have described in great details the training-set generation, hyperparameter optimization and model validation. 
By using the Ni-Pd system as a prototype, we have shown that the JLP achieves strong transferability, it generalizes
from small supercells to significantly larger ones, while requiring only a modest number of training structures. Most
importantly, the JLP is a linear model that depends only on approximately 800 features, thus that its inference 
overheads are modest and the throughput extremely large. 
The JLP is then used to accurately and efficiently predict 
the temperature-dependent Gibbs free energy of formation over an entire range of temperature and composition, 
returning results in close agreement with DFT. This requires 
performing full relaxation for 99,268 configurations and 14,294,592 force calculations, needed to evaluate the 
vibrational contribution to the Gibbs energy.

We have also demonstrated that the same model can be directly used to calculate the elastic tensor, with an error
from the DFT reference of approximately 10\%, enabling the evaluation of the mechanical properties. The framework 
can be further extended to account for thermal expansion so to improve predictions of finite-temperature mechanical, as 
properties. Although here we only considered the experimentally found {\it fcc} phase, the same strategy can be 
pursued to analyse competing crystal system, such as {\it hcp} and {\it bcc}, thereby enabling the construction of 
finite-temperature phase diagrams. 

Beyond disordered binary alloys, the efficiency, transferability, and modest training data requirements of the JLP 
make it particularly well suited for extension to more complex disordered alloys such as HEAs. By enabling accurate 
finite-temperature calculations over large configurational ensembles with modest training data, our framework serves 
as a foundation for systematically exploring thermodynamic and mechanical properties of complex disordered alloys 
and for guiding the design of next-generation materials.

%%%%%%%%%%%%%%%%%%%%%%%%%%%%%%%%%%%%%%%%%%%%%%%%%%%%%%%%%%%%%%%%
%%%%%%%%%%%%%%%%%%%%%%%%%%%%%%%%%%%%%%%%%%%%%%%%%%%%%%%%%%%%%%%%

\section*{CRediT authorship contribution statement}
\textbf{Rutchapon Hunkao:} Conceptualization, Data curation, Formal analysis, Investigation, Methodology, Software, Validation, Visualization, Writing – original draft, Writing – review and editing.
\textbf{Urvesh Patil:} Conceptualization, Methodology, Project administration, Resources, Software, Supervision, Validation.
\textbf{S. Sanvito:} Conceptualization, Funding acquisition, Methodology, Project administration, Resources, Supervision, Validation, Writing – review and editing.

\begin{acknowledgments}
This work has been supported by Science Foundation Ireland through the Advanced Materials and BioEngineering 
Research (AMBER) (Grant: 12/RC/2278$_-$P2). R.H. is grateful for the Thai government scholarship provided 
through the Development and Promotion of Science and Technology Talents Project (DPST).
\end{acknowledgments}

%%%%%%%%%%%%%%%%%%%%%%%%%%%%%%%%%%%%%%%%%%%%%%%%%%%%%%%%%%%%%%%%
%%%%%%%%%%%%%%%%%%%%%%%%%%%%%%%%%%%%%%%%%%%%%%%%%%%%%%%%%%%%%%%%

\appendix

\begin{appendices}

\section{Additional structural information}
\begin{figure}[htp]
    \centering
    \includegraphics[width=0.85\linewidth]{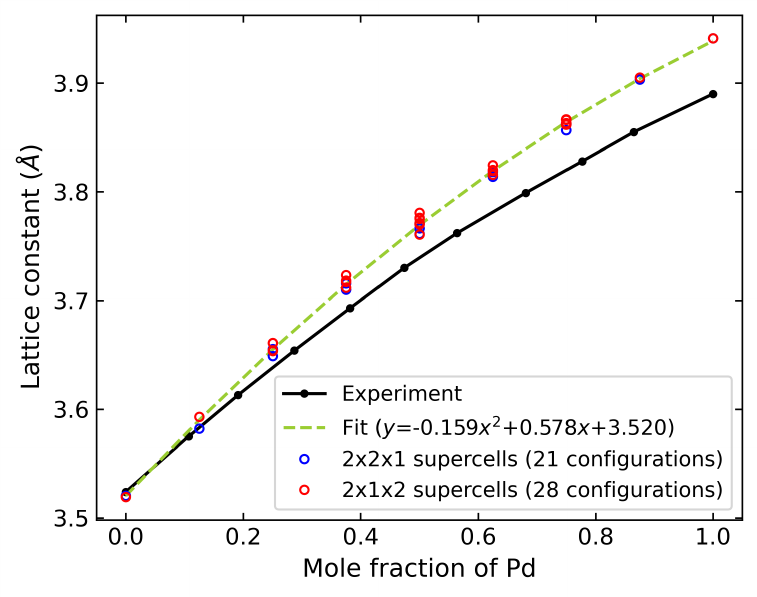}
    \caption{Lattice parameters as a function of the Pd mole fraction. Here we compare results 
    from DFT calculations for relaxed 2$\times$2$\times$1 (blue circles) and 2$\times$1$\times$2 
    (red circles) supercells with the experimental data from \cite{Nash1984}. The best-fit line (green line) 
    is used to calculate the initial lattice parameters (before relaxation) for all the structures contained 
    in the test set.}
    \label{fig:test_initial_lats}
\end{figure}

We report here additional data concerning the structural parameters used for the calculations
presented in the main paper. When generating the initial configurations the lattice parameters 
are set to the reference DFT data. These are obtained by full relaxation of all the unique structures 
constructed from the 2$\times$2$\times$1 and 2$\times$1$\times$2 {\it bct} supercells (21 and 
28 configurations, respectively, covering all stoichiometries) and the results are extended to any 
Pd molar fraction by quadratic interpolation. The best-fit line is obtained by minimizing the errors 
over all configurations. The results of this exercise are presented in Fig.~\ref{fig:test_initial_lats}.

\begin{figure}[htp]
    \centering
    \includegraphics[width=1.0\linewidth]{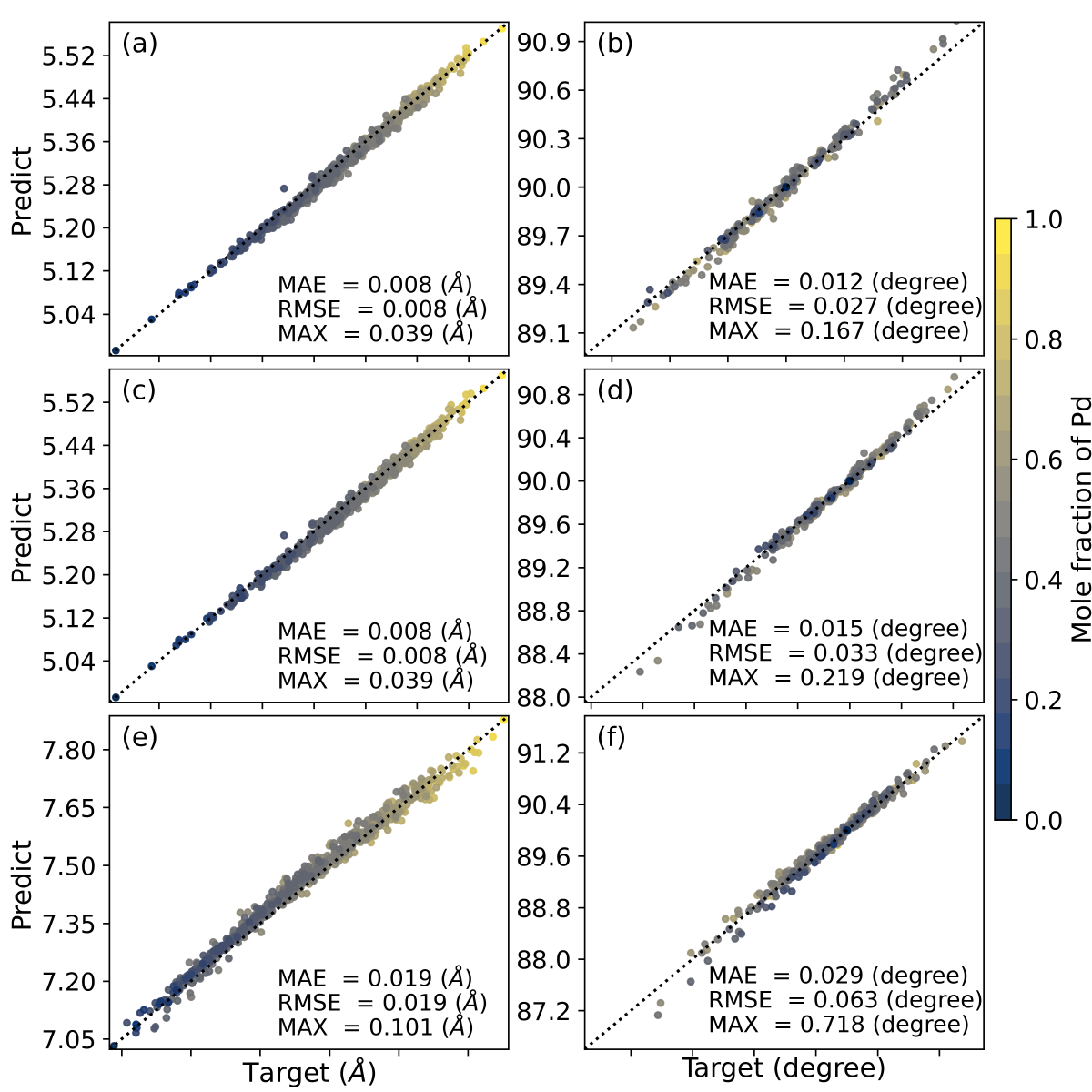}
    \caption{Parity plot for the structural parameters of Ni-Pd alloys, comparing JLP ($y$ axis) predictions 
    against DFT ($x$ axis) reference data. In particular, we present the lattice parameters (a) $a$, (c) $b$ 
    and (e) $c$, and angles (b) $\alpha$, (d) $\beta$ and (f) $\gamma$. Data are for test set {\bf 1}, with the
    dashed line representing the parity line. For each quantity we report the MAE, RMSE and maximum error 
    (MAX).}
    \label{fig:test_lattice_param}
\end{figure}

The equilibrium structural parameters computed with the JLP are compared with the DFT data over the test 
set {\bf 1}, which contains inequivalent 16-atoms supercells. The parity plots for the lattice constants and angles 
associated to such comparison are presented in Fig.~\ref{fig:test_lattice_param}.

\begin{figure}[htp]
    \centering
    \includegraphics[width=1.0\linewidth]{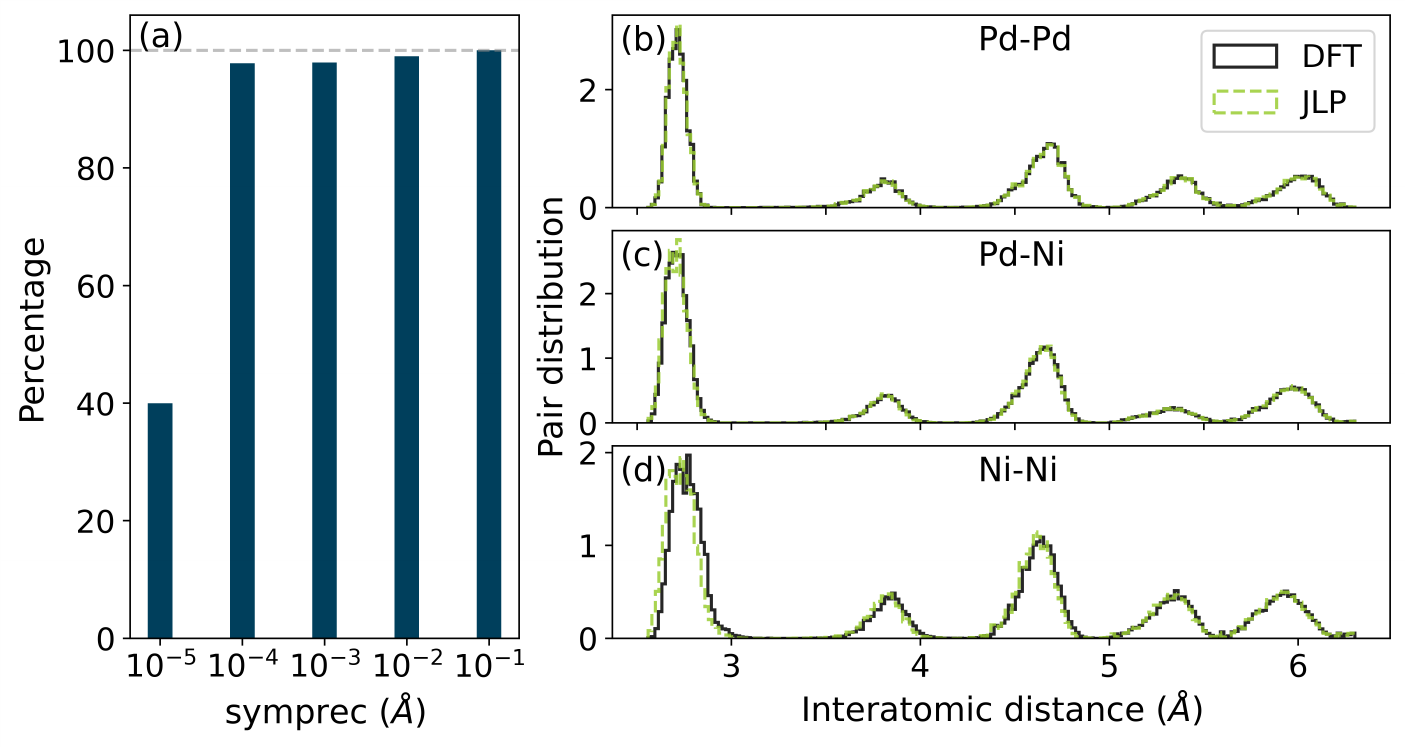}
    \caption{Comparison between the DFT- and JLP-relaxed structures. (a) Percentage of relaxed structures 
    with JLP and DFT that share the same space group symmetry, when calculated with different symmetry 
    tolerance. Pair distribution comparison for JLP- (black lines) and DFT-relaxed (green lines) structures: 
    (a) Pd-Pd, (b) Pd-Ni and (c) Ni-Ni.}
    \label{fig:symprec_pair_dist}
\end{figure}

Finally, in Fig.~\ref{fig:symprec_pair_dist} we compare the final structures relaxed either with DFT or the JLP. 
The symmetry search algorithm \texttt{Spglib} (as implemented in {\sc phonopy}) is used to identify the 
symmetry operations of crystal structures with tolerance \texttt{symprec}. As one can see from panel (a), as the structure tolerance 
threshold is lowered the agreement between the two method increases. Most importantly, at already 
10$^{-4}$~\AA\ the vast majority of structures, 98\%, are classified to share the same space group. This 
indicates that JLP effectively relaxes geometry at a level of precision extremely close to that of DFT.

\section{Effect of the supercell size on the configurational entropy}\label{sec:ideal_entropy}
\begin{figure}
    \centering
    \includegraphics[width=0.9\linewidth]{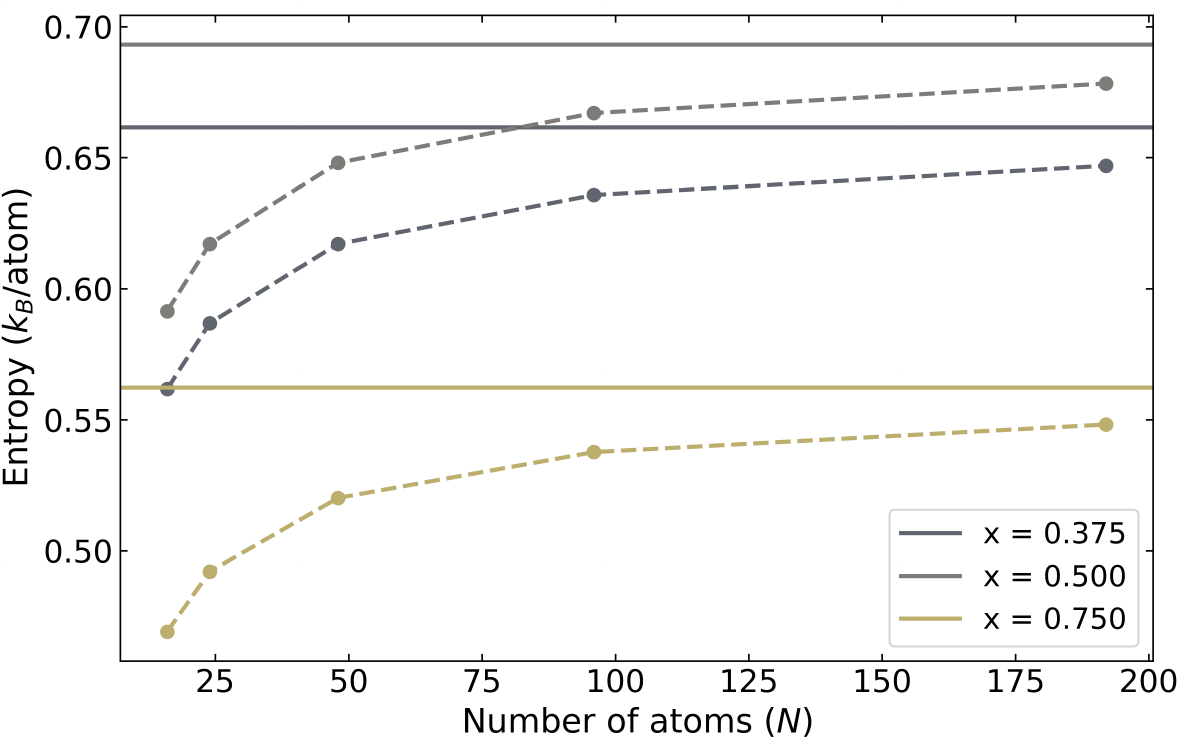}
    \caption{Comparison between the calculated maximum (dashed lines) and the ideal-mixing (solid lines)
    configurational entropies of binary alloys at 37.5\%, 50\%, and 75\% Pd content (indicated by different colors) 
    for different supercell sizes. The maximum entropy is calculated by assuming all configurations are equally 
    probable (high-temperature limit), whereas the ideal-mixing value is given by 
    $S = k_{\mathrm{B}} \sum_i x_i \ln x_i$.}
    \label{fig:max_entropy}
\end{figure}

The comparison between the calculated maximum configurational entropy and the ideal-mixing 
value is shown in Fig.~\ref{fig:max_entropy} for different Pd concentrations as a function of the
number of atoms in the simulation cell. The maximum entropy is reached when all configurations 
are equally probable, and the energy fluctuations between configurations becomes irrelevant. This
is the case at high-temperature, where one should find $S_{\mathrm{max}} = \frac{k_{\mathrm{B}}}{N} \ln \Omega$, 
with $\Omega$ being the total number of available microstates. For example, the number of microstates 
for a half-mixed binary alloy described by a 24-atom cell is $\binom{24}{12} = 2{,}704{,}156$, namely 
$S_{\mathrm{max}} \approx 0.62\ k_{\mathrm{B}}/\text{atom}$. The ideal-mixing entropy is the obtained 
as the limiting value of the maximum configurational entropy when the supercell size approaches infinity,
$S_{\mathrm{ideal}} = -k_{\mathrm{B}} \sum_i x_i \ln x_i$, where $x_i$ is the composition fraction.
From the figure, we observe that the maximum entropy increases with supercell size, but remains always 
below the ideal-mixing value even for cells containing as many as 200 atoms. This behavior reflects the 
finite-size limitation in accessing all possible configurations, as the ideal mixing represents the theoretical 
thermodynamical limit. 
Since the number of microstates grows combinatorially with the supercell size, the enumeration of all 
configurations becomes computationally prohibitive for large cells. Nevertheless, the differences are 
systematic for a given supercell size, as shown in the figure, so that the comparison of different compositions 
using the same supercell size remains meaningful.

\section{Improving the elastic tensor predictions}\label{sec:improve_elastic}
\begin{figure*}[htp]
    \centering
    \includegraphics[width=0.9\linewidth]{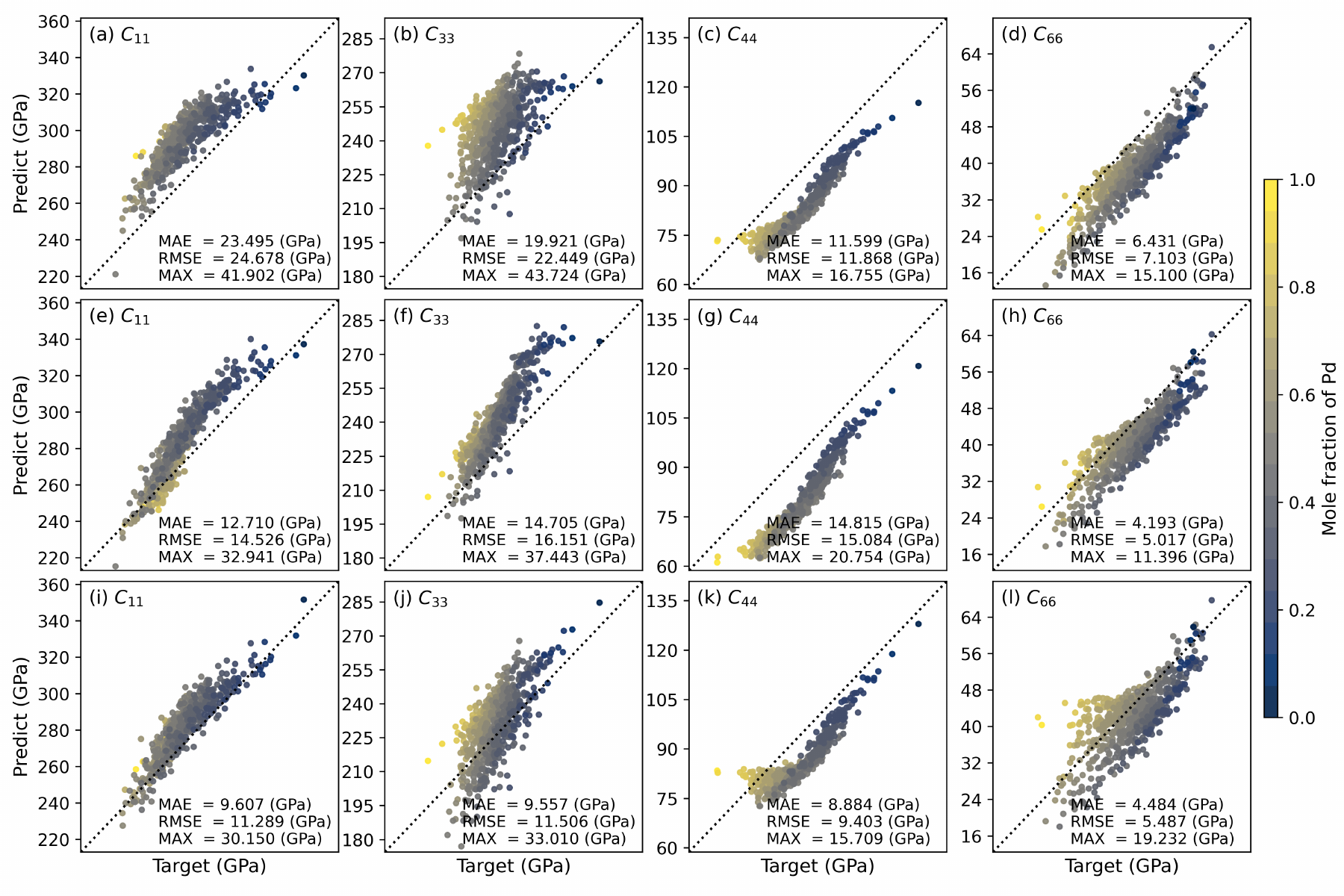}
    \caption{Parity plots for the diagonal components of the elastic constant computed with Model 1, panels (a) through
    (d), the baseline model, panels (e) through (h), and Model 2, panels (i) through (l).}
    \label{fig:elastic_compare}
\end{figure*}

We discuss here how the inclusion of strain-deformed structures in the training data 
affects the accuracy of the elastic-tensor predictions. As summarized in Table \ref{tab:number_strained_cells}, 
three models are compared. Model 1 is trained only on supercells with atomic perturbations; 
Baseline model is the model used throughout this work; Model 2, supplements the perturbed 
supercells with an additional 300 unit cells with both a normal and shear strain of 3\%. 
All models are trained using the the same hyperparameters listed in Table \ref{tab:hyperparam}.
It is clear from the comparison between the baseline model and Model 1 that the stress predictions 
on test set {\bf 1} improve substantially, with the RMSE being reduced from 39.43~meV/atom to 
19.88~meV/atom. This highlights the importance of including strained unit cells for accurate stress 
prediction, a strategy that requires only minimal additional computational costs. A similar trend is 
observed for the elastic tensor predictions, as shown in Fig. \ref{fig:elastic_compare}, where, due 
to cell symmetry, only four components are plotted for comparison.
\begin{table}[htp]
    \centering
    \caption{Number of training structures with different random atomic perturbations ($\delta$), 
    normal ($\varepsilon_{\mu\mu}$) and shear ($\varepsilon_{\mu\nu}$) strains for different models. 
    The RMSE of energies, $E$, forces, $F$, and stress, $W$, and the relaxed energies, 
    $E_\mathrm{relaxed}$, for the test set {\bf 1}, are reported in meV/atom. 
    SC denotes the perturbed supercells (8 atoms), whereas UC denotes the strained unit cells (2 atoms).}
    \begin{tabular}{l|c|c|c}\hline\hline
    & Model 1 & Baseline & Model 2 \\ \hline
SC  $\delta = 3.5\%$ & 731 & 731 & 731 \\
UC $\varepsilon_{\mu\mu}= 3\%$ & - & - & 150 \\
UC $\varepsilon_{\mu\mu}= 5\%$ & - & 328 & - \\
UC $\varepsilon_{\mu\nu, \mu \neq \nu}= 3\%$ & - & - & 150 \\
Total & 731 & 1059 & 1031 \\ \hline
$E$ &                                 1.70 & 1.90   & 1.38 \\
$F$ &                                 9.07 & 10.75 & 10.25 \\
$W$ &                              39.43 & 19.88 & 22.94 \\
$E_\mathrm{relaxed}$ &    1.48 & 1.69   & 1.31 \\ \hline\hline
    \end{tabular}
    \label{tab:number_strained_cells}
\end{table}

\begin{figure}
    \centering
    \includegraphics[width=\linewidth]{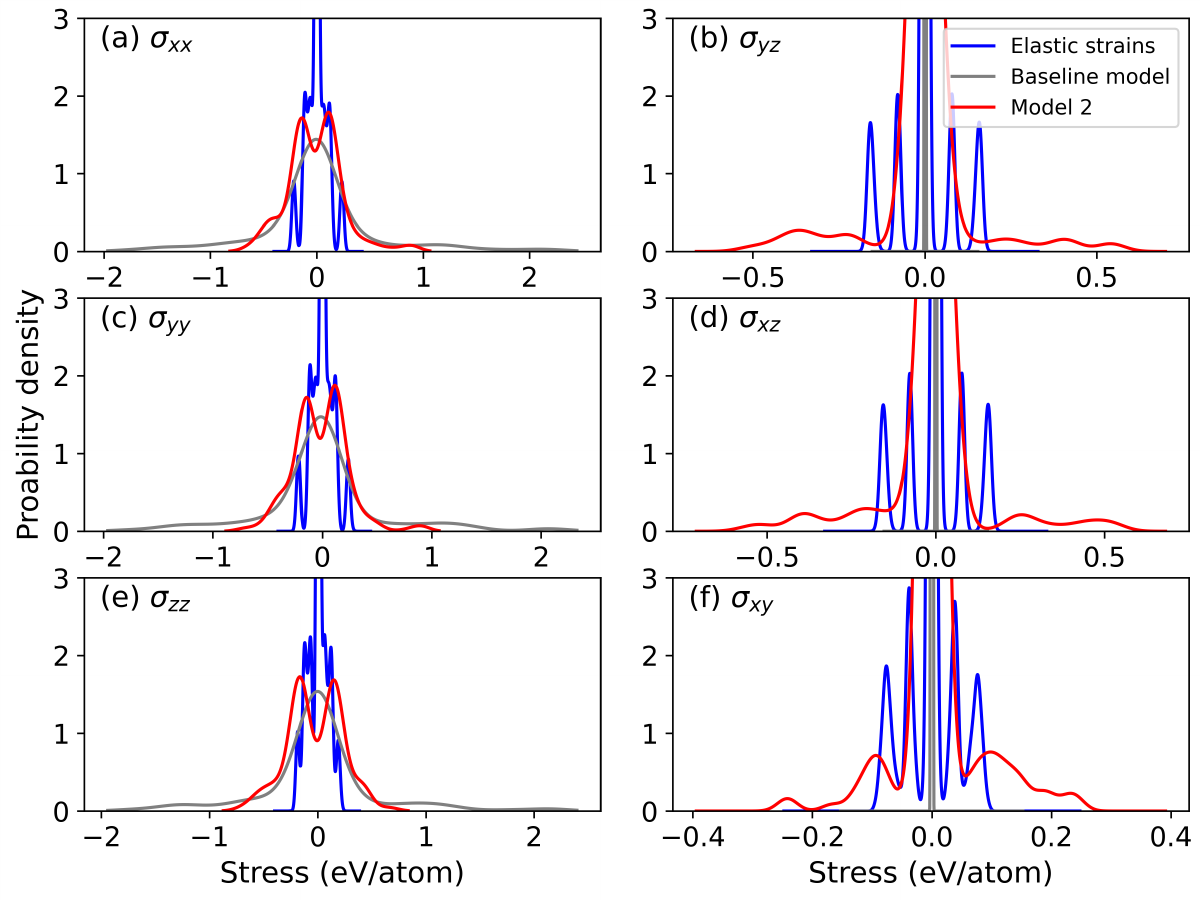}
    \caption{Probability distribution of the stress tensor values for the strain tensor calculation (blue line), and 
    in the strained unit cells as obtained with the baseline model (grey line) and Model 2 (red line). The normal 
    stress components are reported in panels (a) $\sigma_{xx}$, (c) $\sigma_{yy}$, and (e) $\sigma_{zz}$, while 
    the shear stress components are reported in panels (b) $\sigma_{yz}$, (d) $\sigma_{xz}$, and (f) $\sigma_{xy}$.}
    \label{fig:stress_dist}
\end{figure}

To better understand the stress values relevant for the elastic tensor calculations and those included 
in the training sets of different models, their distributions are analyzed. These are presented as probability 
density functions using kernel density estimation. As shown in Fig. \ref{fig:stress_dist}, panels (b), (d) and (f), 
no shear stresses are present in the baseline model, leading to poor predictions of elastic components associated 
with shear stresses, such as $C_{44}$ and $C_{55}$. Furthermore, the stress values in the training 
set span a much broader range than the elastic regime relevant for elastic tensor calculations, as shown 
in Fig. \ref{fig:stress_dist}, panels (a), (c) and (e).
We observe that, by reducing the applied strain from 5\% to 3\% in Model 2, the model achieves improved 
accuracy compared to the baseline model in both energy predictions and structural relaxations, as well as 
more accurate elastic tensor predictions across nearly all components.
\end{appendices}

%%%%%%%%%%%%%%%%%%%%%%%%%%%%%%%%%%%%%%%%%%%%%%%%%%%%%%%%%%%%%%%%
%%%%%%%%%%%%%%%%%%%%%%%%%%%%%%%%%%%%%%%%%%%%%%%%%%%%%%%%%%%%%%%%

\bibliographystyle{apsrev4-2}

\end{document}